\documentclass{aa}  
\pdfoutput=1
\usepackage{natbib}
\usepackage[english]{babel}
\usepackage{graphicx}
\usepackage{txfonts}
\usepackage{url}
\def\ltsima{$\; \buildrel < \over \sim \;$}
\def\lsim{\lower.5ex\hbox{\ltsima}}
\def\gtsima{$\; \buildrel > \over \sim \;$}
\def\gsim{\lower.5ex\hbox{\gtsima}}

\begin{document}

   \title{A time domain experiment with Swift: \\
   monitoring of seven nearby galaxies}

   \author{I. Andreoni\inst{1,2,3}\thanks{ \email{iandreoni@swin.edu.au} }, P. D'Avanzo\inst{1}, S. Campana\inst{1}, M. Branchesi\inst{4,5}, M.G. Bernardini\inst{1}, M. Della Valle\inst{6,7}, F. Mannucci\inst{8},\\ A. Melandri\inst{1}, G. Tagliaferri\inst{1}}

   \institute{INAF, Osservatorio Astronomico di Brera, via E. Bianchi 46, 23807 Merate, Italy
   \and
   Centre for Astrophysics and Supercomputing, Swinburne University of Technology, P.O. Box 218, Hawthorn VIC 3122, Australia    
   \and   
   Dipartimento di Fisica, Universit\`a Degli Studi di Milano, via Celoria 16, 20133 Milano, Italy
   \and
   INFN, Sezione di Firenze, via G. Sansone 1, 50019 Sesto Fiorentino, Italy    
   \and 
   Universit\`a degli Studi di Urbino ``Carlo Bo'', via A. Saffi 2, 61029 Urbino, Italy
   \and
   INAF, Osservatorio Astronomico di Capodimonte, salita Moiariello 16, 80131 Napoli, Italy
   \and
  ICRANET, Piazza della Repubblica 10, 65122, Pescara, Italy
   \and
   INAF, Osservatorio Astrofisico di Arcetri, largo E. Fermi 5, 50125 Firenze, Italy           }

\abstract
   { Focused on the study of transient sources, time domain astronomy is nowadays one of the most active and growing areas of research in astronomy. Most of the present and planned surveys aimed at carrying out time domain studies work in the optical band, and found their searching strategies on fixed cadences. Although nothing similar currently exist in the X--ray and ultraviolet (UV) bands, the \textit{Swift} satellite is certainly the most appropriate available instrument to carry out such surveys.}
   {We aimed to detect a supernova (SN) shock breakout (SBO) in nearby galaxies. The SBO marks the first escape of radiation from the blast wave that breaks through the photosphere of the star and launches the SN ejecta. The detection of a SBO is diagnostic for the radius of the progenitor star, the explosion energy to ejecta mass ratio, and it allows us to determine the onset of the explosion with an accuracy from a few hours to a few seconds.}
   {Using the XRT and UVOT instruments onboard the \textit{Swift} satellite we carried out a weekly cadenced, six months lasting monitoring of seven nearby galaxies, namely NGC\,1084, NGC\,2207/IC\,2163, NGC\,2770, NGC\,4303/M\,61, NGC\,3147, NGC\,3690, NGC\,6754. We searched for variable/transient sources in the collected data. These galaxies have been selected because they are close (distance $\leq 50$\,Mpc), small enough to fit in the \textit{Swift}/UVOT field of view, and host of at least 3 SNe in the last 20\,yr.}
  {We found no evidence for a SN SBO event. Five objects located within the light of the sample galaxies were found to be variable in the X--ray and/or in the UV. These include mainly background AGN and unresolved ULX in NGC 3690. Besides these objects, we found two variable Galactic sources: the known nova CP Draconis (that experienced an outburst during our monitoring) and an uncatalogued eclipsing binary.}
   {Despite the lack of SBO detections, the results of our explorative study encourage the use of \textit{Swift} in further time domain studies. Moreover, since our sample galaxies are within the Universe volume that will be reached by the forthcoming advanced gravitational waves (GW) detectors (a-LIGO/a-Virgo), this work provides an example on how to carry out \textit{Swift} surveys useful to detect the GW signal from SNe, and to detect counterparts to GW triggers.}


\authorrunning{Andreoni et al.}
\titlerunning{A time domain experiment with Swift: monitoring of 7 nearby galaxies}
\maketitle

\section{Introduction}

Time domain astronomy is one of the most active and growing areas of research in astronomy, being able to touch basically every aspect of this science with a different perspective. In the next decade, we expect it to flourish, prompted by facilities like the Large Synoptic Survey Telescope\footnote{lsst.org} in the optical, and the Low-Frequency Array for radio astronomy\footnote{lofar.org} and the Square Kilometre Array\footnote{skatelescope.org} in the radio. Such facilities will revolutionise our understanding of the Universe with nightly searches of large swathes of sky for variable objects and network of robotic telescopes ready to follow-up in greater detail anything of interest. 

Time domain astronomy focuses on transient sources. These might be extragalactic, usually involving catastrophic events (supernovae are the most common examples) or Galactic, usually involving cataclysmic events (novae). In recent years a few single-optical band surveys started exploiting the variable sky, as (among others) the Catalina Sky Survey \citep{Drake2009}, the Palomar Transient Factory \citep{Rau2009}, the PanSTARRS \citep{Stubbs2010}, the La Silla Quest \citep{Rabinowitz2011}, the SUDARE at the VST \citep{Botticella2013} and the SkyMapper \citep{Keller2007} in the 1-m to 2-m telescopes category. These synoptic surveys concentrated on supernovae (SNe), leading, e.g., to the discovery of SN\,2011fe in the Pinwheel galaxy M101 within $\sim$\,1\,day of the explosion \citep{Waagen2011,Piro2014}.

Facilities like GAIA\footnote{sci.esa.int/gaia/} in the optical, the Space Infrared Telescope Facility (\textit{Spitzer}\footnote{spitzer.caltech.edu}) in the infra-red \citep[with projects like SPIRITS;][]{Kasliwal2013}, the \textit{Swift}\footnote{swift.gsfc.nasa.gov}, the Galaxy Evolution Explorer (\textit{GALEX}\footnote{galex.caltech.edu}), and \textit{Fermi}\footnote{fermi.gsfc.nasa.gov} at higher energies have been pioneer of time domain astronomy from space. These open the path to missions such as eROSITA\footnote{mpe.mpg.de/eROSITA} (although more focused on a "static" all-sky survey) and SVOM\footnote{svom.frs}, that will monitor the high energy sky in the near future.

A few thousands of SNe have been discovered so far. These enhanced our understanding of the last stages of massive stellar lives and deaths, and led us to discover that we live in an accelerating Universe that may be dominated by dark energy. In the last few years the decades-ago predicted pulses marking the precise moment that a supernova shock wave breaks out of the progenitor star were discovered \citep{Campana2006, Soderberg2008}. Longer duration (days) shock breakouts (SBOs) have been observed with {\it GALEX} by type\,II SNe related to larger (red giant) progenitors \citep[e.g.,][]{Gezari2008,Schawinski2008}. This is an important tool, since the direct detection of SN progenitors is incredibly difficult, and has only been possible for a small number of nearby core-collapse SNe with pre-explosion high resolution imaging. The {\it Swift} UltraViolet/Optical Telescope \citep[UVOT,][]{Roming2005} is particularly sensitive to the cooling envelope emission, which is bright in the UV for up to several days after the SBO, and can provide estimates of the radius of the progenitor star.

Another ingredient in transient astronomy is the prospect to observe gravitational waves (GWs) in the upcoming years. The second generation ground-based GW detectors are expected to reach sensitivity that will make possible to detect transient GW signals from coalescences of neutron star (NS) and/or stellar-mass black hole binary systems and from core-collapse of massive stars. The advanced LIGO \citep{aLIGO2015} and Virgo \citep{aVirgo2015} detectors in full sensitivity will observe coalescences of NSs up to distance (averaged for sky location and system-orientation) of 200\,Mpc. The core-collapse events are expected to be detectable within a few Mpc \citep{Ott2009,Muller2013,Ott2013, Gossan2015} and up to tens of Mpc for more optimistic models \citep{Fryer2011}.

The present work focuses on SNe, that dominate among transient optical events in nearby galaxies \citep{Rau2009} and aims at detecting the UV/X--ray SBO by monitoring nearby galaxies using \textit{Swift}, that has already proven its ability to redefine time domain astronomy with its instruments \citep{Gehrels2015}. While the UV/X--ray bright SBOs are directly possible electromagnetic (EM) counterparts of the GW signals from core-collapse events, nearby galaxies are in general the host of all the GW transient sources detectable by the GW detectors. This time domain experiment by \textit{Swift} represents an example for a possible monitoring program to detect potential sources of GWs, and at the same time to shed light on UV/X--ray ``transient contaminants'' in galaxy fields. Characterising transient events not directly associated with a GW event will be useful when the EM counterparts of compact object coalescences will be searched in the future.

This paper is organised as follows. In Section\,\ref{sec: target galaxies} we describe our selection criteria for the galaxies and the monitoring characteristics.
In Section\,\ref{sec: analysis methods} we describe the analysis procedures and in Section\,\ref{sec: results} our results. We discuss how well a weekly survey can determine SN SBO onsets in Section\,\ref{sec: SBO} and the role of such a survey in the search for EM counterparts to GW signals in Section\,\ref{sec: GW studies}. 

\section{Target galaxies and monitoring details}\label{sec: target galaxies}
 
Our project aims is to monitor nearby galaxies that are site of an intense production of SNe.
In the selection of the targets one possible problem is dust obscuration. 
Typically high star formation rate galaxies are heavily obscured by dust, making the 
SN detection problematic in the optical. For example, the large number of identified SNe in Arp\,220 (IC\,1127/IC\,4553) came from the radio band and none came from the optical \citep{Lonsdale2006} and \cite{Mannucci2003} showed that the core-collapse SN rate observed at near-IR wavelengths in starburst galaxies is about an order of magnitude larger than in the optical.
For this reason the selection of the sample cannot be based only on the star formation rate.
Hence we adopted a different approach, based on the number of observed SNe in nearby galaxies. Using the Asiago database \citep{Barbon1999}, we selected galaxies in which several SNe were already discovered. 
Requiring at least 3 observed SNe in the last 20 years, we selected 11 galaxies (observed rate $\gsim 0.15$ SN yr$^{-1}$). From these we excluded galaxies in the M\,81/M\,82 group because their separation is larger than the UVOT field of view. Our final selection criteria are:
a) they are close (distance $\lsim 50$\,Mpc, thus within the a-LIGO and a-Virgo horizon) which allows our instruments to resolve their internal structure;
b) their angular size is small enough to fit within the field of view of the \textit{Swift} telescopes (the field of view of the XRT is $23'$ in diameter, or 0.12\,deg$^2$, while the field of view of the UVOT is $17'\times17'$, or 0.08\,deg$^2$).

With our selection criteria the final sample consists of 10 galaxies. {\it Swift} observed 7 of them over the period 2013-2014 (NGC\,5468, NGC\,6946, and NGC\,4038 were left out). The sample is clearly not complete but provides a fair representation of nearby star-forming galaxies. We selected as target galaxies: NGC\,1084, the system NGC\,2207/IC\,2163, NGC\,2770,  NGC\,4303/M\,61, NGC\,3147, NGC\,3690, and NGC\,6754. We present their main features in Table\,\ref{targets_ID} and the details of the observations, such as the dates and the exposure times, in Table\,\ref{observations_NGC1084} to 
\ref{observations_NGC6754}. The monitoring has been carried out on a weekly timescale for about six months. We show examples of UV and X--ray images of the target galaxies in Figs\,\ref{pictures_UV} and \ref{pictures_X}.

\begin{table*}
\centering
\caption{ List of selected targets with their right ascension (RA), declination (DEC), distance, apparent magnitude (V) and angular size. Values provided by the SIMBAD and NED data repository. }
\footnotesize
  \begin{tabular}{c c c c c c}
  \multicolumn{6}{c}{TARGET GALAXIES}\\
    \hline
    \hline
Galaxy	              &RA(J2000)     & Dec(J2000)               &  Dist (Mpc)    & Mag (V)     &Size (arcmin) \\
\hline
NGC1084 	     &02:45:59.926       &$-$07:34:43.10   &  19 	&10.73  &2.62$\times$1.62  \\
NGC2770 	     &09:09:33.622       &$+$33:07:24.29  &  30  &12.80   &2.63$\times$0.69  \\
NGC2207/IC2163&06:16:22.093	 &$-$21:22:21.80   &  40 	&10.65   &4.84$\times$3.29 \\
 NGC\,4303/M\,61   &12:21:54.950       &$+$04:28:24.92  &  15	&9.65     &4.64$\times$3.48 \\
NGC3147   	     &10:16:53.632       &$+$73:24:02.34  &  44	&10.61   &2.85$\times$2.33 \\
NGC3690 	     &11:28:31.326       &$+$58:33:41.80  &  45	&12.86   &1.61$\times$1.41 \\
NGC6754 	     &19:11:25.752       &$-$50:38:31.96   &  42   &11.14   &2.40$\times$0.91 \\
\hline
\end{tabular}
\label{targets_ID}
\end{table*}

\section{Analysis methods}\label{sec: analysis methods} 
 
In our dataset we search for transients, along with increases in luminosity of permanent sources. UV images and X--ray images require different analysis methods, mainly because the angular resolution of the UVOT ($\sim$\,1\,arcsec) is better than the resolution of the XRT ($\sim$\,15\,arcsec) and the density of UV-optical sources largely overcomes the density of the X--ray sources. We used the latest version of the calibration files (CALDB) available in March 2013.

\subsection{XRT data analysis}

We processed the data with {\tt xrtpipeline} v. 0.13.1 to derive calibrated event files. We searched for sources adopting a signal-to-noise ratio threshold of S/N ratio\,=\,3. We took advantage of the HEAdas\footnote{heasarc.nasa.gov/docs/software/lheasoft/} software to identify sources, using the XIMAGE-{\tt detect} tool. First, we searched in every single observation for objects exceeding the threshold. Then, we summed up all the observations of a single galaxy and again we identify all significant sources. 
To build products for sources located outside our target galaxies we took advantage of the \textit{Swift}/XRT online tools\footnote{\url{swift.ac.uk/user_objects/}} \citep{Evans2009}, while for the X--ray sources located inside the target galaxies we carried out a more careful analysis using HEAdas \textit{XSELECT} (using circular apertures, radius\,=20\,arcsec, centred on each identified source).

One X--ray variable source (the nucleus of NGC\,3147; see Section \ref{section: NGC3147}) displayed a count rate high enough to carry out a meaningful X--ray spectral analysis that enabled us to obtain a flux calibrated light curve (Fig.\,\ref{NGC3147_X_center}). Using the \textit{Swift}/XRT online tools we extracted the 0.3-10\,keV spectrum for every single epoch by fitting the data with an absorbed power-law model. The absorption component was fixed to the value of the Galactic column density $N_H$ provided by \cite{Willingale2013}. The results of our analysis are reported in Table \ref{observations_NGC3147}.

In the following, we will consider sources relevant for our purposes if: a) they have been detected above the S/N ratio\,=\,3 threshold in just a single (or a few) epoch; or b) their light curve cannot be satisfactorily fitted with a constant function (with a null hypothesis probability from a constant fit threshold of $p<0.01$).  
 
\subsection{UVOT data analysis} 

We observed with the UVOT/{\it uvm2} filter \citep[central wavelength=2246\,\AA, FWHM=498\,\AA;][]{Poole2008}, which is the `purest' UV filter available (i.e. the filter with the smallest contamination from optical photons). We developed a semi-automatic algorithm based on the image subtraction to search for variable sources in the UVOT data. We took advantage of the ISIS (v. 2.2) \citep[][]{Alard1998, Alard2000} and ESO-ECLIPSE \citep[][]{Devillard1997} packages to perform the image subtraction and to handle the images. We extracted one image per each event file, corresponding to single observations. We aligned the images of a target galaxy taken at different epochs and generated a master image that is the median of all these aligned images. By subtracting this reference frame to every single image, we then searched for positive residuals in the image looking for variable or new sources.
We used the Source Extractor \citep{Bertin2010} to identify those residuals whose flux is greater than 5 times the standard deviation of the local background of the subtracted image. We require the residuals to be symmetric (i.e. related to a point-like source) and with a FWHM comparable to the FWHM of the UVOT (i.e. $\sim$\,5\,arcsec), assuming a symmetrical PSF. 
For each potentially varying sources, we used the HEAdas {\tt uvotproduct} tool to build its UV light curve using apertures with radius\,=\,3\,arcsec.
The count rate to flux ratio conversion was obtained using the UVOT photometric system \citep{Breeveld2011}. No correction for MW reddening was applied in the derived light curves.

\section{Results} \label{sec: results}

In summary, during our monitoring of seven nearby galaxies with the \textit{Swift} XRT and UVOT telescopes we have detected:
\begin{itemize}
\item one variable X--ray source inside the galaxy NGC\,1084 (NGC1084-I1, see Section\,\ref{section: NGC1084}), likely generated by one to three AGNs at higher redshift;
\item one low-luminosity AGN at the centre of the galaxy NGC\,4303 (NGC4303-I1, see Section\,\ref{section: NGC4303}) and one at the centre of NGC\,3147 (NGC3147-I1, see Section\,\ref{section: NGC3147}), both variable in the X--rays but not in the UV band; 
\item one variable X--ray source inside the galaxy NGC\,3690 (possibly due to the unresolved emission of a number of point-like sources detected by {\it Chandra} and positionally coincident with the NGC3690-I1, see Section\,\ref{section: NGC3690});
\item one Seyfert\,1 galaxy located in a region of the sky outside NGC\,4303 (NGC4303-O1, see Section\,\ref{section: NGC4303}) variable in the X--rays as well as in the UV band;
\item one possible quasar in a region of the sky outside the target galaxy NGC\,3147 (NGC3147-O2, see Section\,\ref{section: NGC3147}), variable in the X--rays but outside the field of view of the UVOT; 
\item a Galactic uncatalogued eclipsing binary, located in a region of the sky outside the target galaxy NGC\,2770 (NGC2770-O1, see Section\,\ref{section: NGC2770});
\item a Galactic known nova (CP Draconis) outside the target galaxy NGC\,3147 (NGC3147-O1, see Section\,\ref{section: NGC3147}).

\end{itemize}

We found two false positives in our UVOT dataset, i.e. we identified two sources that survived our selection criteria but that we classified as spurious, ghost images. In one case the ghost lies very close to a bright star, in the other one it is right at the border of the image, where spurious events generated by instrumentation are more likely to happen. 

We discuss now the results of the analysis on each target galaxy.

\subsection{NGC\,1084} \label{section: NGC1084}

\begin{figure}[h]
 \centering
\includegraphics[width=1.\linewidth]{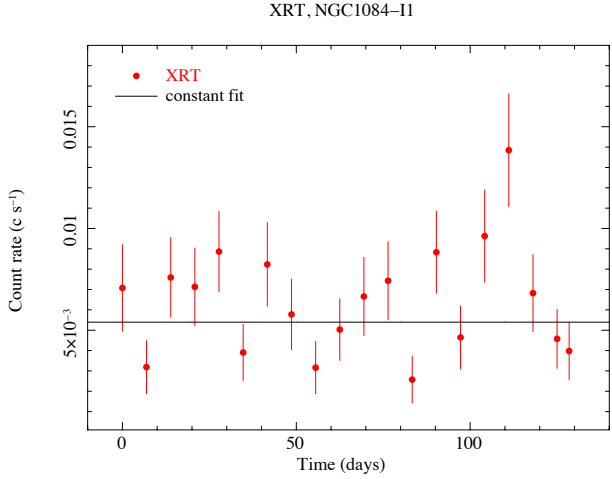}  
\caption{X--ray emission detected from NGC1084-I1. The time zero is set to Mission Elapsed Time (MET)\,=\,405027281\,s (2013-11-01). We extracted this light curve using an aperture with radius 20\,arcsec. The fit performed on the data with a constant function returns a mean count rate of $5.4 \times 10^{-3}$\,counts\,s$^{-1}$. A fit with a constant function gives a $\chi^2$\,=\,40.2 (for 19 degrees of freedom, d.o.f.,), and $p=0.0031$.}
 \label{NGC1084_X_gal}
 \end{figure}

The weekly monitoring of the galaxy NGC\,1084 ranged from the beginning of November 2013 to the second week of March 2014. Good data have been collected for 20 epochs by the XRT and for 19 epochs by the UVOT (see Table\,\ref{observations_NGC1084}).
We found no transient events in the UVOT data. In the XRT data we found that the light curve of an extended region (centred on RA\,=\,02:45:59.9, Dec\,=\,-07:34:23.3, named NGC1048-I1\footnote{All coordinates here and in the following refers to J2000.}) located inside the target galaxy NGC\,1084 varies significantly (Fig.\,\ref{NGC1084_X_gal}). We note however that three candidate active galactic nuclei \citep[AGN,][]{Brough2006, Cavuoti2014} stand within the region, possibly causing the observed variability. We found no significant variability in the UVOT data inside the same region, nor in smaller regions around each AGN. 

\subsection{NGC\,2207/IC\,2163} \label{section: NGC2207}

The weekly monitoring of the system NGC\,2207/IC\,2163 ranged from the beginning of November 2013 to the end of April 2014. Good data have been collected for 27 epochs by the XRT, for 26 epochs by the UVOT (see Table\,\ref{observations_NGC2207}).
We found no transient events in the UVOT data nor in the XRT data.

\subsection{NGC\,2770} \label{section: NGC2770}

\begin{figure}[h]
 \centering
\includegraphics[width=1.\linewidth]{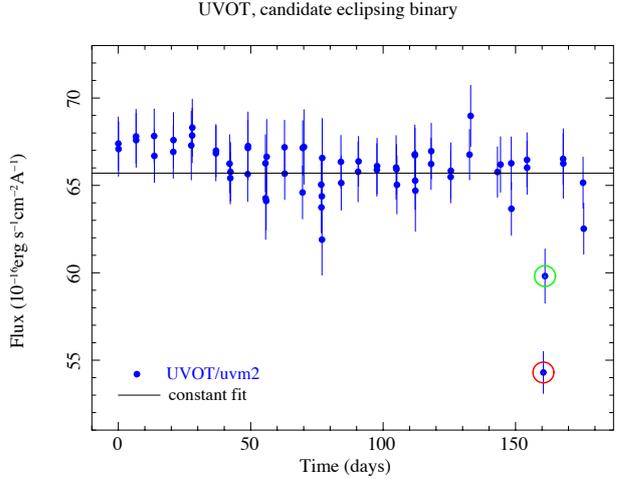}  
 \caption{UVOT/\textit{uvm2} light curve of the candidate eclipsing binary system NGC2770-O1. The time zero is set to MET\,=\,408372298\,s (2013-12-10). The fit performed on the data with a constant function returns a mean flux of $6.6\times10^{-15}$\,erg\,s$^{-1}$\,cm$^{-2}$\,A$^{-1}$. The fit provides a $\chi^2$\,=\,147.1 (63 d.o.f.) and a probability  $p=9\times 10^{-11}$. The flux suddenly decreases by the $\sim 20\%$ (red circle) and the $\sim 10 \%$ (green circle) with respect to the constant value. Both points were collected on May 20\textsuperscript{th} 2014, in two different orbits of the \textit{Swift} satellite. We found no X--ray counterpart to this source. } 
 \label{NGC2770_ecl_binary_UV}
 \end{figure}

The weekly monitoring of the galaxy NGC\,2770 ranged from the second week of December 2013 to the beginning of June 2014. Good data have been collected for 25 epochs by both the XRT and the UVOT (see Table\,\ref{observations_NGC2770}).
In the UVOT data we found no transient events that could be interesting to our purposes. However, we identified a sudden drop of the UV flux emitted by a star outside the target galaxy. This source is located at coordinates RA\,=\,9:09:34.8, Dec\,=\,+33:09:28.4 (NGC2770-O1). Its light curve keeps a constant trend till the flux suddenly decreases in two consecutive points (both occurring on May 20\textsuperscript{th} 2014, as they correspond to two different orbits of the \textit{Swift} satellite) by the $\sim$\,20$\%$ and the $\sim$\,10\,$\%$ with respect to the mean value (Fig.\,\ref{NGC2770_ecl_binary_UV}). 
This source is classified as an \textit{r}\,=\,12\,mag star in the SDSS. We have not detected any X--ray counterpart to this source in our XRT dataset. We suggest that it may be an eclipsing binary system. 
In the XRT data we found no transient events and no variability inside or outside the galaxy NGC\,2770. 

\subsection{ NGC\,4303/M\,61} \label{section: NGC4303}

\begin{figure}[h]
 \centering
\includegraphics[width=1.\linewidth]{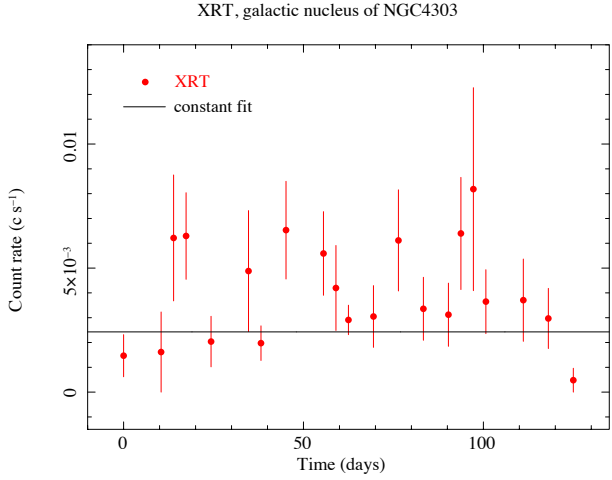}  
 \caption{X--ray light curve of the galactic nucleus of NGC\,4303. The time zero is set to MET\,=\,412217321\,s (2014-01-24). The fit performed on the data with a constant function returns a mean rate of $2.4\times 10^{-3}$\,counts\,s$^{-1}$ ($\chi^2=47.30$ for 20 d.o.f. and $p=0.0005$). No significant UV variability is present in our dataset.} 
 \label{NGC4303_X_gal}
 \end{figure}

\begin{figure}[h]
 \centering
\includegraphics[width=1.\linewidth]{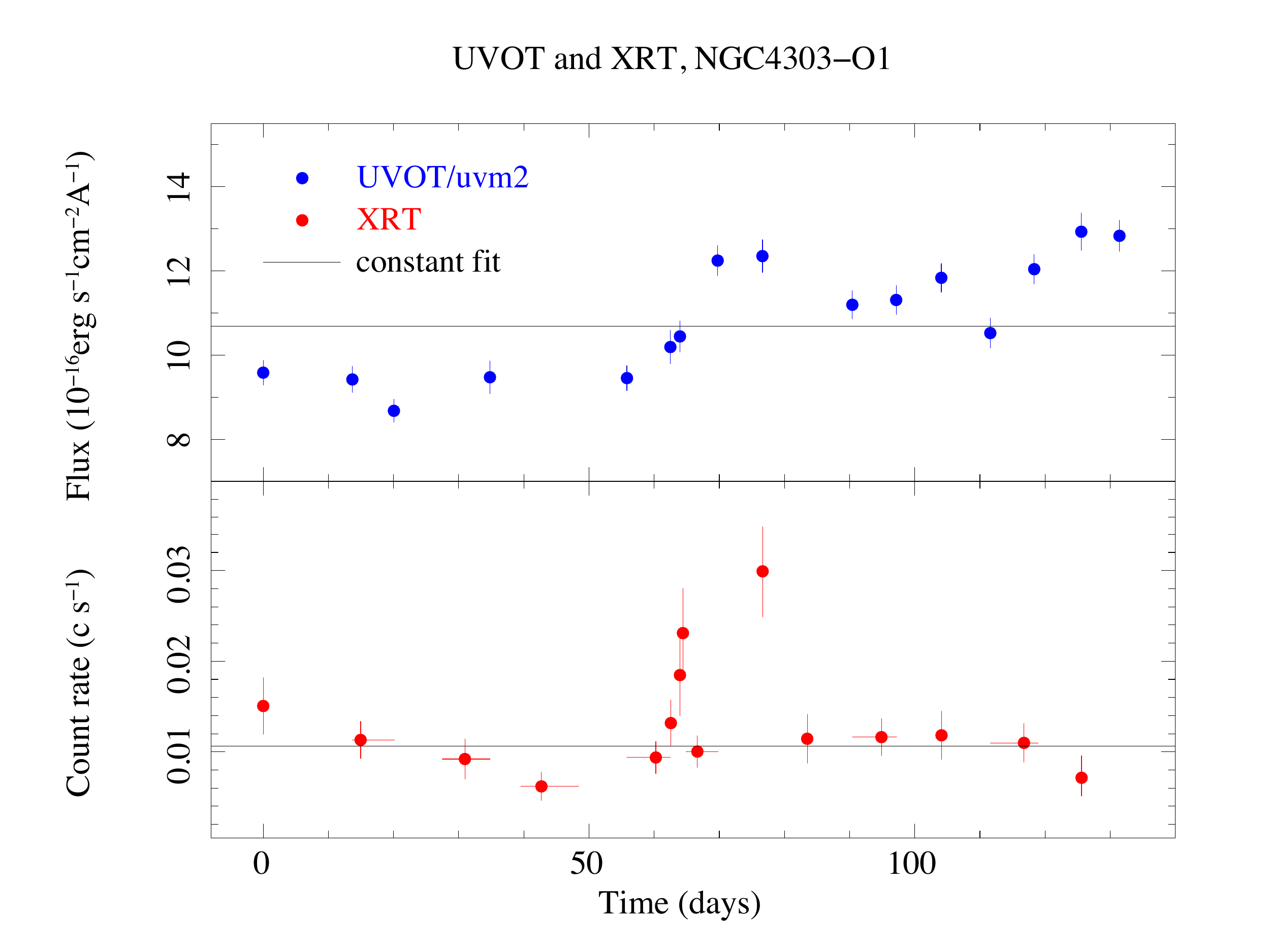}  
\caption{UVOT/\textit{uvm2} and XRT light curves of the NGC4303-O1, identified as a Seyfert\,1 galaxy (catalogued as 2XMM\,J122137.8+043025). The time zero is set to MET\,=\,412217321\,s (2014-01-24). The fit performed on the data with a constant function
returns a mean UV flux (upper panel, aperture with radius\,=\,5\,arcsec) of 1.1\,$\times\,10^{-15}$\,erg\,s$^{-1}$\,cm$^{-2}$\,A$^{-1}$, $\chi^2=242.8$ (15 d.o.f.), and a null probability.
The fit with a constant function on the X--ray data (lower panel) returns a mean rate of 1.06\,$\times 10^{-2}$\,counts\,s$^{-1}$,
$\chi^2=39.15$ (14 d.o.f.), and  $p=0.0003$.}
  \label{NGC4303_X_sg}
 \end{figure}

The weekly monitoring of the galaxy NGC\,4303 - better known as M\,61 - ranged from the final week of January 2014 to the first week of June 2014. Good data have been collected for 21 epochs by the XRT, for 20 epochs by the UVOT (see Table\,\ref{observations_NGC4303}). This galaxy hosts a low luminosity AGN \citep{Brinkmann1994}.
We detected no transient events in the UVOT data. The X--ray emission of the active nucleus is significantly variable, as expected (Fig.\,\ref{NGC4303_X_gal}). The flux of its UV counterpart 
does not vary significantly in the same time frame. As for the other galaxies we analysed the X--ray sources located outside the target galaxy. One of them (NGC4303-O1) shows significant variability in its light curve (Fig.\,\ref{NGC4303_X_sg}). The source is coincident with  a known Seyfert\,1 Galaxy \citep[RA\,=\,12:21:38.0, Dec\,=\,+04:30:26.4,][]{Spinelli2006}. Looking at the UVOT data, we found that its counterpart at lower energy is highly variable, too (Fig.\,\ref{NGC4303_X_sg}). This UV variability has not been detected by our algorithm during the UVOT data analysis because we optimised the procedure for point-like sources inside nearby galaxies. In this case the variable emission comes from the unresolved nucleus of an extended source, which leads the image subtraction to produce an asymmetrical residual that was discarded by our automatic procedure.

On October 2014 the Type\,Ia supernova SN\,2014dt exploded in M\,61 \citep{Nakano2014, Ochner2014}. We monitored this galaxy with the \textit{Swift} satellite till the first week of June 2014 (Table\,\ref{observations_NGC4303}). Our dataset lacks of any precursor signal that can relate to SN\,2014dt, both in the X--rays and in the UV band. We also found no evidence of any luminous progenitor, in agreement with \cite{Foley2015}.

\subsection{NGC\,3147} \label{section: NGC3147}

\begin{figure}[h]
 \centering
  \includegraphics[width=1\linewidth]{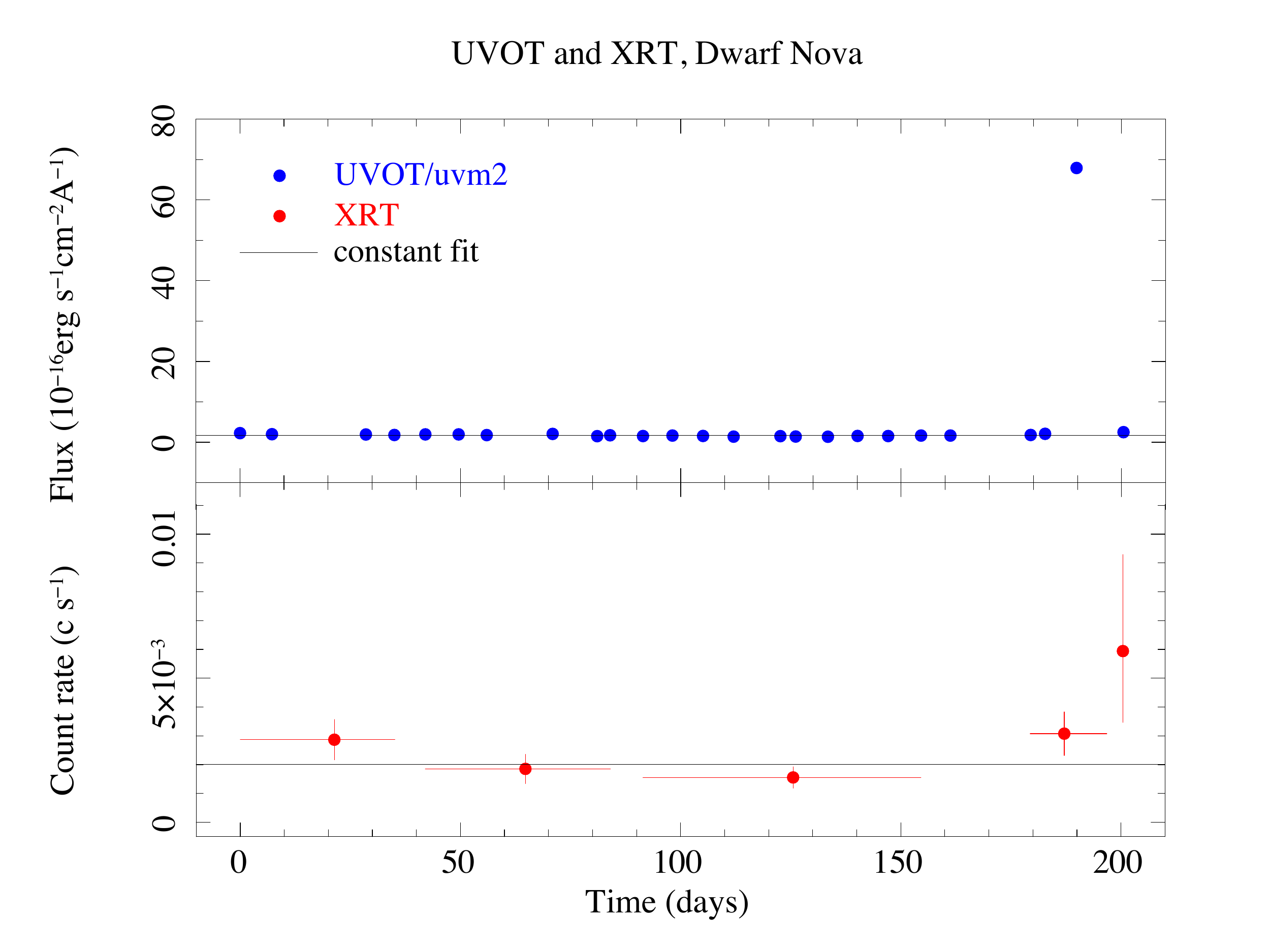}  
  \caption{ UVOT/\textit{uvm2} light curve (upper panel) extracted with an aperture (radius\,=\,6\,arcsec) centred on the dwarf nova CP Draconis (NGC3147-O1). The time zero is set to MET\,=\,409627840\,s (2013-12-25). The fit with a constant function provides a $\chi^2=3685$ (25 d.o.f.), and a null probability. The X--ray counterpart to this source (lower panel) does not show any significant variability, as the fit with a constant function returns a $\chi^2=6.426$ (4 d.o.f.) and a probability $p=0.1695$.}
  \label{NGC3147_UV_nova}
 \end{figure}

The weekly monitoring of NGC\,3147, a Seyfert\,2 galaxy hosting an AGN \citep{Ptak1996}, ranged from the Christmas day 2013 to the second week of July 2014. Good data have been collected for 27 epochs by both the XRT and the UVOT (see Table\,\ref{observations_NGC3147}).
We have detected one transient event (a known dwarf nova, located outside the target galaxy) in our UVOT dataset, at coordinates RA\,=\,10:15:39.8, Dec\,=\,+73:26:05.0. We identified the outburst of the dwarf nova CP Draconis (NGC3147-O1) in the observation of July 2\textsuperscript{nd}, 2014. The outburst activity of this object is known to be recurrent \citep{Shears2011}. The UV light curve (Fig.\,\ref{NGC3147_UV_nova}, upper panel) shows that the flux rises up to at least ($6.79 \pm 0.13$)$\,\times\,10^{-15}$\,erg\,s$^{-1}$\,cm$^{-2}$\,A$^{-1}$ during the outburst, while it lies around $\sim$\,$1.4 \times\,10^{-16}$\,erg\,s$^{-1}$\,cm$^{-2}$\,A$^{-1}$ during the quiescent phase. The X--ray counterpart to this source (Fig.\,\ref{NGC3147_UV_nova}, lower panel) consists of a weak signal that shows no significant variability during our monitoring. 

\begin{figure}[h]
 \centering
  \includegraphics[width=1.\linewidth]{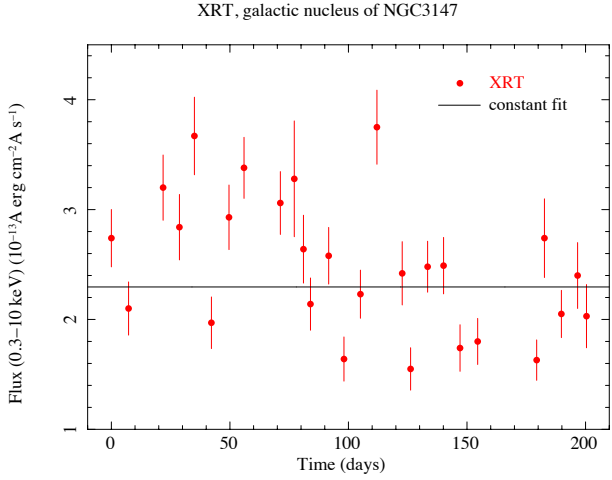}  
  \caption{X--ray light curve of the active galactic nucleus \citep{Ptak1996} of NGC\,3147 (NGC3147-I1). We detected enough photons from this source to build a spectrum at each epoch and convert the count rate to flux. The spectral indexes found at each epoch are listed in Table\,\ref{observations_NGC3147}. The time zero is set to MET\,=\,409627840\,s (2013-12-25). The mean flux is $2.3 \times 10^{-13}$\,erg\,cm$^{-2}$\,s$^{-1}$. The fit with a constant gives $\chi^2=141.6$ (26 d.o.f.) and a null probability. The UV counterpart of this source is not significantly variable. }
  \label{NGC3147_X_center}
 \end{figure}

In the XRT data we found two variable sources. The first, as expected, is the active nucleus of the NGC\,3147 (Fig.\,\ref{NGC3147_X_center}), NGC3147-I1. Its variability in the UV band is however not significant. The second variable X--ray source, NGC3147-O2, is a candidate quasar \citep[RA\,=\,10:14:21.5, Dec\,=\,+73:17:26.0,][]{Flesch2010}, located outside the target galaxy, but too far from NGC\,3147 to fit within the UVOT field of view.

\begin{figure}[h]
 \centering
  \includegraphics[width=1.\linewidth]{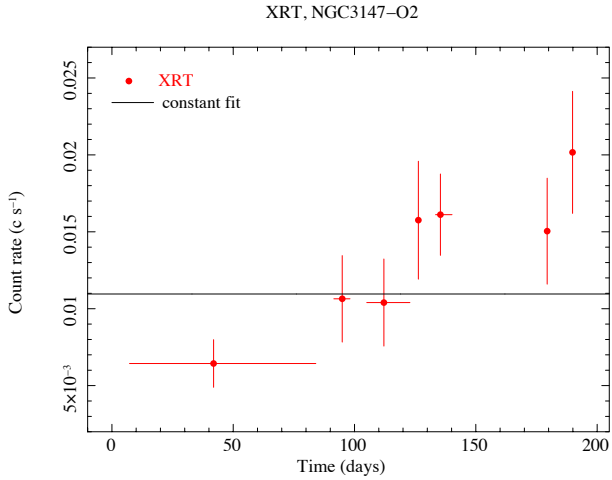}  
  \caption{XRT light curve of the candidate quasar \citep{Flesch2010} (NGC\,3147-O2). The time zero is set to MET\,=\,409627840\,s (2013-12-25). The mean rate is 1.10\,$\times 10^{-2}$\,counts\,s$^{-1}$, $\chi^2=20.73$ (6 d.o.f.) and $p=0.0021$. }
  \label{NGC3147_X_source}
 \end{figure}

\subsection{NGC\,3690} \label{section: NGC3690}
\begin{figure}[h]
 \centering
  \includegraphics[width=1.\linewidth]{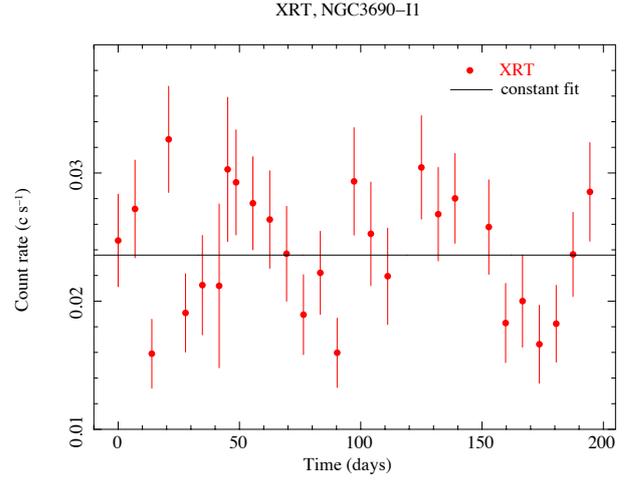}  
  \caption{X--ray radiation detected inside the target galaxy NGC\,3690 (NGC3690-I1). This galaxy is not characterised by an active nucleus, so the radiation is likely generated by a number of X--ray emitters inside the galaxy. The time zero is set to MET\,=\,411013065\,s (2014-01-10). The fit performed on the data with a constant function returns a rate of $2.36\times 10^{-2}$\,counts\,s$^{-1}$, a $\chi^2$\,=\,62.85 (35 d.o.f.) and $p=0.0026$.}
  \label{NGC3690_X_center}
 \end{figure}

The weekly monitoring of the galaxy NGC\,3690 ranged from the second week of January 2014 to the last week of September 2014. Good data have been collected for 35 by the XRT and for 34 epochs by the UVOT (see Table\,\ref{observations_NGC3690}).
We identified no transient events in the UVOT data collected during our observations. In the XRT data we identified a powerful and significantly variable emission generated inside the target galaxy NGC\,3690 at coordinates RA\,=\,11:28:31.7, Dec\,=\,+58:33:46.4 (NGC3690-I1, see Fig.\,\ref{NGC3690_X_center}). Several X--ray sources are present within NGC\,3690 as observed by \textit{Chandra} \citep{Zezas2003} and \textit{NuSTAR} \citep{Ptak2015} at a position consistent with our variable X--ray source. Two of them have been identified as AGNs, some others might be classified as ultra-luminous X--ray sources (ULX). Considering the angular resolution of the XRT, we cannot definitely state which source(s) is responsible for the observed variability.

\subsection{NGC\,6754} \label{section: NGC6754}

The weekly monitoring of the galaxy NGC\,6754 ranged from the first week of April 2014 to the end of September 2014. Good data have been collected for 27 epochs by both the XRT and the UVOT (see Table\,\ref{observations_NGC6754}).
We found no transient events in the UVOT data, as well as in the XRT data. 

\section{Supernova shock breakouts} \label{sec: SBO}

The SBO, that is the soft X--ray and UV outburst expected at the birth of SNe \citep{Colgate1974, Falk1978, Klein1978, Matzner1999}, marks the first escape of radiation when the blast wave breaks through the surface of the star and launches the SN ejecta. Its detection can enable an early follow-up of the SN, with a detailed study of the luminosity and temperature evolution of the early thermal expansion phase. This information can be used as a diagnostic for the radius of the progenitor star, and the explosion energy to the ejecta mass ratio \citep{Waxman2007}.

The short duration (seconds to hours) of SBOs and the lack of sensitive wide-field searches at high energies make their discovery very hard. However, 
the \textit{Swift}/XRT and UVOT provided evidences of their existence by detecting the early X--ray/UV emission a few days before the SNe 2006aj associated with the GRB\,060218 \citep{Campana2006} and before the SNe 2008D \citep{Soderberg2008}. Also the \textit{GALEX} satellite detected serendipitous early UV events associated to the SNe SNLS-04D2dc \citep{Schawinski2008, Gezari2008}, SNLS-06D1jd \citep{Gezari2008} and 2010aq \citep{Gezari2010}.

These few observations demonstrate the capability to constrain the onset time ($T_0$) of SBOs, which  marks the moment of the star explosion.
For red supergiant progenitors (radius $\sim$\,$10^{13}$\,cm, then Type\,II-P SNe), $T_0$ was estimated with 0.75\,days accuracy for SNLS-06D1jd and 0.35\,days for SNLS-04D2dc \citep{Gezari2008}, while \cite{Schawinski2008} could estimate it with $\sim$\,1\,hour precision. A 0.02\,days accuracy was reached for SN\,2010aq \citep{Gezari2010} by fitting the data with theoretical models \citep{Rabinak2011}. For Wolf-Rayet star progenitors (radius $\sim$\,$10^{11}$\,cm, then Type\,Ibc SNe), the onset of the X--ray SBO associated with SN\,2008D \citep{Soderberg2008} was determined with $9^{+8}_{-20}$\,s accuracy. A precise determination of the ($T_0$), achieved with SBO observations and accurate theoretical models describing how the shock wave propagates through the exploding star,  is fundamental in the multi-messenger studies, which use SBOs as triggers to search for GWs and neutrinos (see Section\,\ref{sec: GW studies}). 

The effectiveness of our survey to detect an SBO was estimated to be 15.9$^{+11,1}_{-6.6}\%$ from Type\,II-P SNe, of 0.03$\%$ ($\pm$\,0.02$\%$) from Type\,Ibc SNe. To calculate these probabilities we assumed a (conservative) detection rate of 3 SNe in the last 20\,y (0.15\,SNe\,yr$^{-1}$) for each galaxy of our sample \citep{Barbon1999}, with a 90$\%$ interval corresponding to a Poissonian mean rate of (0.08-0.27) SNe\,yr$^{-1}$. We considered the UV emission lasting 3.5\,days and 1000\,s for Type II-P and Tipe\,Ibc SNe, respectively. We also accounted for the 65$\%$ of SNe exploding as Type II-P, the 25$\%$ exploding as Type Ibc \citep{Smartt2009, vandenBergh2005, Cappellaro1999}.

\section{Counterparts to gravitational waves} \label{sec: GW studies}

The multi-frequency periodic monitoring of nearby galaxies presented in this paper represents a useful explorative study to test the capabilities of the \textit{Swift} satellite for joint surveys with the GW detectors. Our target, the SBO, is an expected EM counterpart to GW signals from core-collapse SNe, and  our sample galaxies are within the Universe volume ($<$ 200\,Mpc) that will be reached by the forthcoming advanced GW detectors (a-LIGO and a-Virgo) for binary coalescing NS systems.

Two types of multi-messenger searches can benefit from this type of \textit{Swift} survey: the search that uses external EM trigger events to search for gravitational wave events \citep[e.g.][]{Abadie2012a} and the EM follow up of GW candidate events \citep[e.g.][]{Abadie2012b}. 

The {\it Swift} detection of a transient phenomenon that is expected to be a GW emitter, like local (within few tens of Mpc) SNe or ``orphan'' gamma-ray bursts \citep{Ghirlanda2015}, provides  timing and sky position that can be used in the search for the GW signal. The use of external triggers improves significantly  
the GW search sensitivity \citep{Was2012} with respect to all-sky  searches. The probability to detect SNe during a {\it Swift} monitoring like the one of the present work does not significantly increase the total number of local SNe detected by optical surveys that can be used as external triggers. 
However, the X--ray/UV detection can give a more precise way to estimate the explosion time $T_0$ and the time window ($T_0$ uncertainties) where to search for the GW signal. 
As discussed in the previous section, the detection of a SBO associated with accurate theoretical models permits to estimate the time of the core-collapse (and thus of the GW emission) with precision from hours down to seconds. On the other hand, the detection of a ``canonical'' SN or gamma-ray burst afterglow, even with a frequent (daily) monitoring, will likely not provide an estimate of $T_0$ with precision better than 1 day.

The narrow field of the {\it Swift}/XRT and UVOT represents a 
limitation in the search for the EM counterpart to a GW source given that, even in the era of the advanced detectors, the sky localization uncertainty of GW signals is expected to be large \citep[hundreds of square degrees,][]{Singer2014, Aasi2013}. For this reason, the effective strategy for the \textit{Swift} satellite  is to observe a limited number of fields inside the error region, namely those containing known, nearby galaxies 
\citep{Evans2012, Kanner2012, Evans2015, Gehrels2015}. 

Any search for EM counterparts to GW signals has to face the problem of contamination from variable/transient objects, unrelated with the GW source, detected within its localization region \citep[e.g.,][]{Cowperthwaite2015}. 
In our six-months monitoring (weekly cadenced) of seven closeby galaxies, we found that the main contaminant (variable) sources in both X--rays (limiting flux\,$\sim 1.3 \times 10^{-13}$\,erg\,s$^{-1}$\,cm$^{-2}$) and UV (limiting flux\,$\sim 1.5 \times 10^{-13}$\,erg\,s$^{-1}$\,cm$^{-2}$) are AGNs (NGC1084-I1, NGC4303-I1, NGC3147-I1) and likely ULX (NGC3690-I1), if the search is limited within the galaxies. For these events, the use of astronomical catalogues (mainly through the Vizier database\footnote{vizier.u-strasbg.fr/viz-bin/VizieR}) was sufficient to securely identify most of them, both in X--rays and in UV. This suggests that an automated and rapid use of archival data will be extremely useful to reduce the contaminants. 

In our survey of 7 galaxies covering a sky area 0.014\,deg$^2$, no other contaminant brighter than $3 \times 10^{-12}$\,erg\,s$^{-1}$\,cm$^{-2}$ (0.2-2\,KeV band) was found. \cite{Kanner2013} estimated $4 \times 10^{-4}$ transients per square degree with a flux brighter than the above threshold, spatially coincident with galaxies, and after rejecting previously identified AGNs. According to Poissonian statistics, the probability to observe zero transients in our small observed area is approximately 1.

\section{Conclusions} \label{sec: conclusion} 

We have observed seven nearby galaxies with a weekly cadence for about six months over the period 2013-2014 with the \textit{Swift}/UVOT and XRT. Aiming at the detection of a SN SBO, we chose to target nearby galaxies with high rate of SNe recently detected in the optical ($\ge 3$\,SN\ in the last 20 years), namely: NGC\,1084, the system NGC\,2207/IC\,2163, NGC\,2770,  NGC\,4303/M\,61, NGC\,3147, NGC\,3690, NGC\,6754. We estimated the probability to detect a SBO (including the cooling phase) during our survey to be 15.9$^{+11,1}_{-6.6}\%$ for Type\,II-P SNe, 0.03$\%$ ($\pm$\,0.02$\%$) for Type\,Ibc SNe. We detected several variable sources in both UVOT and XRT data, but we relate none of them to a SN SBO event.

This work demonstrates that the \textit{Swift} satellite is suitable to carry on targeted, multi-wavelength, time domain observations. 
We discussed the importance of the detection of X--ray/UV SBOs, that are difficult to detect but can be a powerful tool to gain information about the progenitor star and the properties of the ejecta.
Also, the onset of the SBO is the best EM mark of the moment of the explosion of massive stars. While the accuracy on the onset time provided by the study of “canonical” optical SNe and GRB afterglows cannot be better than about 1 day, for the X--ray/UV SBO it can be as accurate as a few hours down to seconds.

We stress that a survey like the one we carried out with \textit{Swift} is and will be useful in the framework of multi-messenger astronomy. It can significantly contribute to the detection of EM signature of the GW sources. Indeed, SNe are putative GW emitters, and the detection of their SBO would increase our ability to constrain the time window in which the explosion takes place, which is crucial to search for GW signals. Besides this, SNe occurring in dense circumstellar environment can be particularly bright in the X-rays \citep{Chandra2012} with an emission that can be observable for weeks/months, providing a useful probe for the circumstellar matter properties and for the final stages of the progenitor star \citep{Ofek2013}.

Finally, all galaxies included in our sample are close enough ($<$50\,Mpc) to lay within the a-LIGO and a-Virgo horizon for NS-NS and black hole-NS merger events. Despite the limited field of view of \textit{Swift}/XRT and UVOT, choosing nearby galaxies as target of a regular, periodic observation is a winning strategy to trigger the GW signal search, to find possible EM counterparts during follow-up of GW detections \citep[see, e.g.,][]{Evans2015} and to rule out contaminants by the study of their light curves.    

\begin{acknowledgements}

GT, SC, PDA, and AM acknowledge the support from ASI I/004/11/0. MB acknowledges financial support from the Italian Ministry of Education, University and Research (MIUR) through grant FIRB 2012 RBFR12PM1F. MGB acknowledges the T-Rex project.

\end{acknowledgements}

\bibliographystyle{aa} 
\bibliography{/Users/iandreoni/Desktop/References/references}

\begin{thebibliography}{68}
\expandafter\ifx\csname natexlab\endcsname\relax\def\natexlab#1{#1}\fi

\bibitem[{{Abadie} {et~al.}(2012){Abadie}, {Abbott}, {Abbott}, {Abbott},
  {AberNathy}, {Accadia}, {Acernese}, {Adams}, {Adhikari}, {Affeldt}, \&
  et~al.}]{Abadie2012a}
{Abadie}, J., {Abbott}, B.~P., {Abbott}, R., {et~al.} 2012, ApJ, 760, 12

\bibitem[{{Acernese} {et~al.}(2015){Acernese}, {Agathos}, {Agatsuma}, {Aisa},
  {Allemandou}, {Allocca}, {Amarni}, {Astone}, {Balestri}, {Ballardin}, \&
  et~al.}]{aVirgo2015}
{Acernese}, F., {Agathos}, M., {Agatsuma}, K., {et~al.} 2015, Classical and
  Quantum Gravity, 32, 024001

\bibitem[{{Alard}(2000)}]{Alard2000}
{Alard}, C. 2000, A\&AS, 144, 363

\bibitem[{{Alard} \& {Lupton}(1998)}]{Alard1998}
{Alard}, C. \& {Lupton}, R.~H. 1998, ApJ, 503, 325

\bibitem[{{Barbon} {et~al.}(1999){Barbon}, {Buond{\'{\i}}}, {Cappellaro}, \&
  {Turatto}}]{Barbon1999}
{Barbon}, R., {Buond{\'{\i}}}, V., {Cappellaro}, E., \& {Turatto}, M. 1999,
  A\&A, 139, 531

\bibitem[{{Bertin} \& {Arnouts}(2010)}]{Bertin2010}
{Bertin}, E. \& {Arnouts}, S. 2010, {SExtractor: Source Extractor},
  Astrophysics Source Code Library

\bibitem[{{Botticella} {et~al.}(2013){Botticella}, {Cappellaro}, {Pignata},
  {Baruffolo}, {Benetti}, {Bufano}, {Capaccioli}, {Cascone}, {Covone}, {Della
  Valle}, {Grado}, {Greggio}, {Limatola}, {Paolillo}, {Pastorello},
  {Tomasella}, {Turatto}, \& {Vaccari}}]{Botticella2013}
{Botticella}, M.~T., {Cappellaro}, E., {Pignata}, G., {et~al.} 2013, The
  Messenger, 151, 29

\bibitem[{{Breeveld} {et~al.}(2011){Breeveld}, {Landsman}, {Holland}, {Roming},
  {Kuin}, \& {Page}}]{Breeveld2011}
{Breeveld}, A.~A., {Landsman}, W., {Holland}, S.~T., {et~al.} 2011, in American
  Institute of Physics Conference Series, Vol. 1358, American Institute of
  Physics Conference Series, ed. J.~E. {McEnery}, J.~L. {Racusin}, \&
  N.~{Gehrels}, 373--376

\bibitem[{{Brinkmann} {et~al.}(1994){Brinkmann}, {Siebert}, \&
  {Boller}}]{Brinkmann1994}
{Brinkmann}, W., {Siebert}, J., \& {Boller}, T. 1994, A\&A, 281, 355

\bibitem[{{Brough} {et~al.}(2006){Brough}, {Forbes}, {Kilborn}, {Couch}, \&
  {Colless}}]{Brough2006}
{Brough}, S., {Forbes}, D.~A., {Kilborn}, V.~A., {Couch}, W., \& {Colless}, M.
  2006, MNRAS, 369, 1351

\bibitem[{{Campana} {et~al.}(2006){Campana}, {Mangano}, {Blustin}, {Brown},
  {Burrows}, {Chincarini}, {Cummings}, {Cusumano}, {Della Valle}, {Malesani},
  {M{\'e}sz{\'a}ros}, {Nousek}, {Page}, {Sakamoto}, {Waxman}, {Zhang}, {Dai},
  {Gehrels}, {Immler}, {Marshall}, {Mason}, {Moretti}, {O'Brien}, {Osborne},
  {Page}, {Romano}, {Roming}, {Tagliaferri}, {Cominsky}, {Giommi}, {Godet},
  {Kennea}, {Krimm}, {Angelini}, {Barthelmy}, {Boyd}, {Palmer}, {Wells}, \&
  {White}}]{Campana2006}
{Campana}, S., {Mangano}, V., {Blustin}, A.~J., {et~al.} 2006, Nat, 442, 1008

\bibitem[{{Cappellaro} {et~al.}(1999){Cappellaro}, {Evans}, \&
  {Turatto}}]{Cappellaro1999}
{Cappellaro}, E., {Evans}, R., \& {Turatto}, M. 1999, A\&A, 351, 459

\bibitem[{{Cavuoti} {et~al.}(2014){Cavuoti}, {Brescia}, {D'Abrusco}, {Longo},
  \& {Paolillo}}]{Cavuoti2014}
{Cavuoti}, S., {Brescia}, M., {D'Abrusco}, R., {Longo}, G., \& {Paolillo}, M.
  2014, MNRAS, 437, 968

\bibitem[{{Chandra} {et~al.}(2012){Chandra}, {Chevalier}, {Irwin}, {Chugai},
  {Fransson}, \& {Soderberg}}]{Chandra2012}
{Chandra}, P., {Chevalier}, R.~A., {Irwin}, C.~M., {et~al.} 2012, ApJ, 750, L2

\bibitem[{{Chevalier} \& {Klein}(1978)}]{Klein1978}
{Chevalier}, R.~A. \& {Klein}, R.~I. 1978, ApJ, 219, 994

\bibitem[{{Colgate}(1974)}]{Colgate1974}
{Colgate}, S.~A. 1974, ApJ, 187, 333

\bibitem[{{Cowperthwaite} \& {Berger}(2015)}]{Cowperthwaite2015}
{Cowperthwaite}, P.~S. \& {Berger}, E. 2015, ArXiv e-prints 1503.07869

\bibitem[{{Devillard}(1997)}]{Devillard1997}
{Devillard}, N. 1997, The Messenger, 87, 19

\bibitem[{{Drake} {et~al.}(2009){Drake}, {Djorgovski}, {Mahabal}, {Beshore},
  {Larson}, {Graham}, {Williams}, {Christensen}, {Catelan}, {Boattini},
  {Gibbs}, {Hill}, \& {Kowalski}}]{Drake2009}
{Drake}, A.~J., {Djorgovski}, S.~G., {Mahabal}, A., {et~al.} 2009, ApJ, 696,
  870

\bibitem[{{Evans} {et~al.}(2009){Evans}, {Beardmore}, {Page}, {Osborne},
  {O'Brien}, {Willingale}, {Starling}, {Burrows}, {Godet}, {Vetere}, {Racusin},
  {Goad}, {Wiersema}, {Angelini}, {Capalbi}, {Chincarini}, {Gehrels}, {Kennea},
  {Margutti}, {Morris}, {Mountford}, {Pagani}, {Perri}, {Romano}, \&
  {Tanvir}}]{Evans2009}
{Evans}, P.~A., {Beardmore}, A.~P., {Page}, K.~L., {et~al.} 2009, MNRAS, 397,
  1177

\bibitem[{{Evans} {et~al.}(2012){Evans}, {Fridriksson}, {Gehrels}, {Homan},
  {Osborne}, {Siegel}, {Beardmore}, {Handbauer}, {Gelbord}, {Kennea}, \&
  et~al.}]{Evans2012}
{Evans}, P.~A., {Fridriksson}, J.~K., {Gehrels}, N., {et~al.} 2012, ApJS, 203,
  28

\bibitem[{{Evans} {et~al.}(2015){Evans}, {Osborne}, {Kennea}, {Campana},
  {O'Brien}, {Tanvir}, {Racusin}, {Burrows}, {Cenko}, \& {Gehrels}}]{Evans2015}
{Evans}, P.~A., {Osborne}, J.~P., {Kennea}, J.~A., {et~al.} 2015, ArXiv e-print
  1506.01624

\bibitem[{{Falk}(1978)}]{Falk1978}
{Falk}, S.~W. 1978, ApJ, 225, L133

\bibitem[{{Flesch}(2010)}]{Flesch2010}
{Flesch}, E. 2010, PASA, 27, 283

\bibitem[{{Foley} {et~al.}(2015){Foley}, {Van Dyk}, {Jha}, {Clubb},
  {Filippenko}, {Mauerhan}, {Miller}, \& {Smith}}]{Foley2015}
{Foley}, R.~J., {Van Dyk}, S.~D., {Jha}, S.~W., {et~al.} 2015, ApJ, 798, L37

\bibitem[{{Fryer} \& {New}(2011)}]{Fryer2011}
{Fryer}, C.~L. \& {New}, K.~C.~B. 2011, Living Reviews in Relativity, 14, 1

\bibitem[{{Gehrels} \& {Cannizzo}(2015)}]{Gehrels2015}
{Gehrels}, N. \& {Cannizzo}, J.~K. 2015, Journal of High Energy Astrophysics,
  7, 2

\bibitem[{{Gezari} {et~al.}(2008){Gezari}, {Dessart}, {Basa}, {Martin},
  {Neill}, {Woosley}, {Hillier}, {Bazin}, {Forster}, {Friedman}, {Le Du},
  {Mazure}, {Morrissey}, {Neff}, {Schiminovich}, \& {Wyder}}]{Gezari2008}
{Gezari}, S., {Dessart}, L., {Basa}, S., {et~al.} 2008, ApJ, 683, L131

\bibitem[{{Gezari} {et~al.}(2010){Gezari}, {Rest}, {Huber}, {Narayan},
  {Forster}, {Neill}, {Martin}, {Valenti}, {Smartt}, {Chornock}, {Berger},
  {Soderberg}, {Mattila}, {Kankare}, {Burgett}, {Chambers}, {Dombeck}, {Grav},
  {Heasley}, {Hodapp}, {Jedicke}, {Kaiser}, {Kudritzki}, {Luppino}, {Lupton},
  {Magnier}, {Monet}, {Morgan}, {Onaka}, {Price}, {Rhoads}, {Siegmund},
  {Stubbs}, {Tonry}, {Wainscoat}, {Waterson}, \& {Wynn-Williams}}]{Gezari2010}
{Gezari}, S., {Rest}, A., {Huber}, M.~E., {et~al.} 2010, ApJ, 720, L77

\bibitem[{{Ghirlanda} {et~al.}(2015){Ghirlanda}, {Salvaterra}, {Campana},
  {Vergani}, {Japelj}, {Bernardini}, {Burlon}, {D'Avanzo}, {Melandri},
  {Gomboc}, {Nappo}, {Paladini}, {Pescalli}, {Salafia}, \&
  {Tagliaferri}}]{Ghirlanda2015}
{Ghirlanda}, G., {Salvaterra}, R., {Campana}, S., {et~al.} 2015, A\&A, 578, A71

\bibitem[{{Gossan} {et~al.}(2015){Gossan}, {Sutton}, {Stuver}, {Zanolin},
  {Gill}, \& {Ott}}]{Gossan2015}
{Gossan}, S.~E., {Sutton}, P., {Stuver}, A., {et~al.} 2015, ArXiv e-prints
  1511.02836

\bibitem[{{Kanner} {et~al.}(2013){Kanner}, {Baker}, {Blackburn}, {Camp},
  {Mooley}, {Mushotzky}, \& {Ptak}}]{Kanner2013}
{Kanner}, J., {Baker}, J., {Blackburn}, L., {et~al.} 2013, ApJ, 774, 63

\bibitem[{{Kanner} {et~al.}(2012){Kanner}, {Camp}, {Racusin}, {Gehrels}, \&
  {White}}]{Kanner2012}
{Kanner}, J., {Camp}, J., {Racusin}, J., {Gehrels}, N., \& {White}, D. 2012,
  ApJ, 759, 22

\bibitem[{{Kasliwal} {et~al.}(2013){Kasliwal}, {Cao}, {Surace}, {Helou},
  {Williams}, {Kulkarni}, {Smith}, {Armus}, {Bond}, {Cantiello}, {Gehrz},
  {Kobulnicky}, {Langer}, {Levesque}, {Masci}, {Mohamed}, {Ofek},
  {Parthasarathy}, {Tang}, {van Dyk}, \& {Whitelock}}]{Kasliwal2013}
{Kasliwal}, M., {Cao}, Y., {Surace}, J., {et~al.} 2013, {SPIRITS: SPitzer
  InfraRed Intensive Transients Survey}, Spitzer Proposal

\bibitem[{{Keller} {et~al.}(2007){Keller}, {Schmidt}, {Bessell}, {Conroy},
  {Francis}, {Granlund}, {Kowald}, {Oates}, {Martin-Jones}, {Preston},
  {Tisserand}, {Vaccarella}, \& {Waterson}}]{Keller2007}
{Keller}, S.~C., {Schmidt}, B.~P., {Bessell}, M.~S., {et~al.} 2007, PASA, 24, 1

\bibitem[{{LIGO Scientific Collaboration} {et~al.}(2013){LIGO Scientific
  Collaboration}, {Virgo Collaboration}, {Aasi}, {Abadie}, {Abbott}, {Abbott},
  {Abbott}, {Abernathy}, {Accadia}, {Acernese}, \& et~al.}]{Aasi2013}
{LIGO Scientific Collaboration}, {Virgo Collaboration}, {Aasi}, J., {et~al.}
  2013, ArXiv e-prints 1304.0670

\bibitem[{{LIGO Scientific Collaboration} {et~al.}(2012){LIGO Scientific
  Collaboration}, {Virgo Collaboration}, {Abadie}, {Abbott}, {Abbott},
  {Abbott}, {Abernathy}, {Accadia}, {Acernese}, {Adams}, \&
  et~al.}]{Abadie2012b}
{LIGO Scientific Collaboration}, {Virgo Collaboration}, {Abadie}, J., {et~al.}
  2012, A\&A, 539, A124

\bibitem[{{Lonsdale} {et~al.}(2006){Lonsdale}, {Diamond}, {Thrall}, {Smith}, \&
  {Lonsdale}}]{Lonsdale2006}
{Lonsdale}, C.~J., {Diamond}, P.~J., {Thrall}, H., {Smith}, H.~E., \&
  {Lonsdale}, C.~J. 2006, ApJ, 647, 185

\bibitem[{{Mannucci} {et~al.}(2003){Mannucci}, {Maiolino}, {Cresci}, {Della
  Valle}, {Vanzi}, {Ghinassi}, {Ivanov}, {Nagar}, \&
  {Alonso-Herrero}}]{Mannucci2003}
{Mannucci}, F., {Maiolino}, R., {Cresci}, G., {et~al.} 2003, A\&A, 401, 519

\bibitem[{{Matzner} \& {McKee}(1999)}]{Matzner1999}
{Matzner}, C.~D. \& {McKee}, C.~F. 1999, ApJ, 510, 379

\bibitem[{{M{\"u}ller} {et~al.}(2013){M{\"u}ller}, {Janka}, \&
  {Marek}}]{Muller2013}
{M{\"u}ller}, B., {Janka}, H.-T., \& {Marek}, A. 2013, ApJ, 766, 43

\bibitem[{{Nakano} {et~al.}(2014){Nakano}, {Itagaki}, {Guido}, {Nicolini},
  {Howes}, {Kiyota}, {Masi}, {Catalano}, {Vagnozzi}, \& {Munari}}]{Nakano2014}
{Nakano}, S., {Itagaki}, K., {Guido}, E., {et~al.} 2014, CBET, 4011, 1

\bibitem[{{Ochner} {et~al.}(2014){Ochner}, {Tomasella}, {Benetti},
  {Cappellaro}, {Elias-Rosa}, {Pastorello}, \& {Turatto}}]{Ochner2014}
{Ochner}, P., {Tomasella}, L., {Benetti}, S., {et~al.} 2014, ATel, 6648, 1

\bibitem[{{Ofek} {et~al.}(2013){Ofek}, {Fox}, {Cenko}, {Sullivan}, {Gnat},
  {Frail}, {Horesh}, {Corsi}, {Quimby}, {Gehrels}, {Kulkarni}, {Gal-Yam},
  {Nugent}, {Yaron}, {Filippenko}, {Kasliwal}, {Bildsten}, {Bloom},
  {Poznanski}, {Arcavi}, {Laher}, {Levitan}, {Sesar}, \& {Surace}}]{Ofek2013}
{Ofek}, E.~O., {Fox}, D., {Cenko}, S.~B., {et~al.} 2013, ApJ, 763, 42

\bibitem[{{Ott}(2009)}]{Ott2009}
{Ott}, C.~D. 2009, Classical and Quantum Gravity, 26, 063001

\bibitem[{{Ott} {et~al.}(2013){Ott}, {Abdikamalov}, {M{\"o}sta}, {Haas},
  {Drasco}, {O'Connor}, {Reisswig}, {Meakin}, \& {Schnetter}}]{Ott2013}
{Ott}, C.~D., {Abdikamalov}, E., {M{\"o}sta}, P., {et~al.} 2013, ApJ, 768, 115

\bibitem[{{Piro} \& {Nakar}(2014)}]{Piro2014}
{Piro}, A.~L. \& {Nakar}, E. 2014, ApJ, 784, 85

\bibitem[{{Poole} {et~al.}(2008){Poole}, {Breeveld}, {Page}, {Landsman},
  {Holland}, {Roming}, {Kuin}, {Brown}, {Gronwall}, {Hunsberger}, {Koch},
  {Mason}, {Schady}, {vanden Berk}, {Blustin}, {Boyd}, {Broos}, {Carter},
  {Chester}, {Cucchiara}, {Hancock}, {Huckle}, {Immler}, {Ivanushkina},
  {Kennedy}, {Marshall}, {Morgan}, {Pandey}, {de Pasquale}, {Smith}, \&
  {Still}}]{Poole2008}
{Poole}, T.~S., {Breeveld}, A.~A., {Page}, M.~J., {et~al.} 2008, MNRAS, 383,
  627

\bibitem[{{Ptak} {et~al.}(2015){Ptak}, {Hornschemeier}, {Zezas}, {Lehmer},
  {Yukita}, {Wik}, {Antoniou}, {Argo}, {Ballo}, {Bechtol}, {Boggs}, {Della
  Ceca}, {Christensen}, {Craig}, {Hailey}, {Harrison}, {Krivonos}, {Maccarone},
  {Stern}, {Tatum}, {Venters}, \& {Zhang}}]{Ptak2015}
{Ptak}, A., {Hornschemeier}, A., {Zezas}, A., {et~al.} 2015, ApJ, 800, 104

\bibitem[{{Ptak} {et~al.}(1996){Ptak}, {Yaqoob}, {Serlemitsos}, {Kunieda}, \&
  {Terashima}}]{Ptak1996}
{Ptak}, A., {Yaqoob}, T., {Serlemitsos}, P.~J., {Kunieda}, H., \& {Terashima},
  Y. 1996, ApJ, 459, 542

\bibitem[{{Rabinak} \& {Waxman}(2011)}]{Rabinak2011}
{Rabinak}, I. \& {Waxman}, E. 2011, ApJ, 728, 63

\bibitem[{{Rabinowitz} {et~al.}(2011){Rabinowitz}, {Tourtellotte}, {Baltay},
  {Bailyn}, {Coppi}, {Rojo}, {Folatelli}, \& {Hoyer}}]{Rabinowitz2011}
{Rabinowitz}, D.~L., {Tourtellotte}, S., {Baltay}, C., {et~al.} 2011, in
  Bulletin of the American Astronomical Society, Vol.~43, American Astronomical
  Society Meeting Abstracts 217, 126.04

\bibitem[{{Rau} {et~al.}(2009){Rau}, {Kulkarni}, {Law}, {Bloom}, {Ciardi},
  {Djorgovski}, {Fox}, {Gal-Yam}, {Grillmair}, {Kasliwal}, {Nugent}, {Ofek},
  {Quimby}, {Reach}, {Shara}, {Bildsten}, {Cenko}, {Drake}, {Filippenko},
  {Helfand}, {Helou}, {Howell}, {Poznanski}, \& {Sullivan}}]{Rau2009}
{Rau}, A., {Kulkarni}, S.~R., {Law}, N.~M., {et~al.} 2009, PASP, 121, 1334

\bibitem[{{Roming} {et~al.}(2005){Roming}, {Kennedy}, {Mason}, {Nousek}, {Ahr},
  {Bingham}, {Broos}, {Carter}, {Hancock}, {Huckle}, {Hunsberger}, {Kawakami},
  {Killough}, {Koch}, {McLelland}, {Smith}, {Smith}, {Soto}, {Boyd},
  {Breeveld}, {Holland}, {Ivanushkina}, {Pryzby}, {Still}, \&
  {Stock}}]{Roming2005}
{Roming}, P.~W.~A., {Kennedy}, T.~E., {Mason}, K.~O., {et~al.} 2005, SSRv, 120,
  95

\bibitem[{{Schawinski} {et~al.}(2008){Schawinski}, {Justham}, {Wolf},
  {Podsiadlowski}, {Sullivan}, {Steenbrugge}, {Bell}, {R{\"o}ser}, {Walker},
  {Astier}, {Balam}, {Balland}, {Carlberg}, {Conley}, {Fouchez}, {Guy},
  {Hardin}, {Hook}, {Howell}, {Pain}, {Perrett}, {Pritchet}, {Regnault}, \&
  {Yi}}]{Schawinski2008}
{Schawinski}, K., {Justham}, S., {Wolf}, C., {et~al.} 2008, Science, 321, 223

\bibitem[{{Shears} {et~al.}(2011){Shears}, {Boyd}, {Buczynski}, {Martin},
  {Messier}, {Miller}, {Oksanen}, {Pietz}, {Skillman}, {Staels}, \&
  {Vanmunster}}]{Shears2011}
{Shears}, J., {Boyd}, D., {Buczynski}, D., {et~al.} 2011, Journal of the
  British Astronomical Association, 121, 36

\bibitem[{{Singer} {et~al.}(2014){Singer}, {Price}, {Farr}, {Urban}, {Pankow},
  {Vitale}, {Veitch}, {Farr}, {Hanna}, {Cannon}, {Downes}, {Graff}, {Haster},
  {Mandel}, {Sidery}, \& {Vecchio}}]{Singer2014}
{Singer}, L.~P., {Price}, L.~R., {Farr}, B., {et~al.} 2014, ApJ, 795, 105

\bibitem[{{Smartt}(2009)}]{Smartt2009}
{Smartt}, S.~J. 2009, ARA\&A, 47, 63

\bibitem[{{Soderberg} {et~al.}(2008){Soderberg}, {Berger}, {Page}, {Schady},
  {Parrent}, {Pooley}, {Wang}, {Ofek}, {Cucchiara}, {Rau}, {Waxman}, {Simon},
  {Bock}, {Milne}, {Page}, {Barentine}, {Barthelmy}, {Beardmore}, {Bietenholz},
  {Brown}, {Burrows}, {Burrows}, {Byrngelson}, {Cenko}, {Chandra}, {Cummings},
  {Fox}, {Gal-Yam}, {Gehrels}, {Immler}, {Kasliwal}, {Kong}, {Krimm},
  {Kulkarni}, {Maccarone}, {M{\'e}sz{\'a}ros}, {Nakar}, {O'Brien}, {Overzier},
  {de Pasquale}, {Racusin}, {Rea}, \& {York}}]{Soderberg2008}
{Soderberg}, A.~M., {Berger}, E., {Page}, K.~L., {et~al.} 2008, Nat, 453, 469

\bibitem[{{Spinelli} {et~al.}(2006){Spinelli}, {Storchi-Bergmann}, {Brandt}, \&
  {Calzetti}}]{Spinelli2006}
{Spinelli}, P.~F., {Storchi-Bergmann}, T., {Brandt}, C.~H., \& {Calzetti}, D.
  2006, ApJS, 166, 498

\bibitem[{{Stubbs} {et~al.}(2010){Stubbs}, {Doherty}, {Cramer}, {Narayan},
  {Brown}, {Lykke}, {Woodward}, \& {Tonry}}]{Stubbs2010}
{Stubbs}, C.~W., {Doherty}, P., {Cramer}, C., {et~al.} 2010, ApJS, 191, 376

\bibitem[{{The LIGO Scientific Collaboration} {et~al.}(2015){The LIGO
  Scientific Collaboration}, {Aasi}, {Abbott}, {Abbott}, {Abbott}, {Abernathy},
  {Ackley}, {Adams}, {Adams}, {Addesso}, \& et~al.}]{aLIGO2015}
{The LIGO Scientific Collaboration}, {Aasi}, J., {Abbott}, B.~P., {et~al.}
  2015, Classical and Quantum Gravity, 32, 074001

\bibitem[{{van den Bergh} {et~al.}(2005){van den Bergh}, {Li}, \&
  {Filippenko}}]{vandenBergh2005}
{van den Bergh}, S., {Li}, W., \& {Filippenko}, A.~V. 2005, PASP, 117, 773

\bibitem[{{Waagen}(2011)}]{Waagen2011}
{Waagen}, E.~O. 2011, AAVSO Alert Notice, 446, 1

\bibitem[{{Waxman} {et~al.}(2007){Waxman}, {M{\'e}sz{\'a}ros}, \&
  {Campana}}]{Waxman2007}
{Waxman}, E., {M{\'e}sz{\'a}ros}, P., \& {Campana}, S. 2007, ApJ, 667, 351

\bibitem[{{W{\c a}s} {et~al.}(2012){W{\c a}s}, {Sutton}, {Jones}, \&
  {Leonor}}]{Was2012}
{W{\c a}s}, M., {Sutton}, P.~J., {Jones}, G., \& {Leonor}, I. 2012, Phys. Rev.
  D., 86, 022003

\bibitem[{{Willingale} {et~al.}(2013){Willingale}, {Starling}, {Beardmore},
  {Tanvir}, \& {O'Brien}}]{Willingale2013}
{Willingale}, R., {Starling}, R.~L.~C., {Beardmore}, A.~P., {Tanvir}, N.~R., \&
  {O'Brien}, P.~T. 2013, MNRAS, 431, 394

\bibitem[{{Zezas} {et~al.}(2003){Zezas}, {Ward}, \& {Murray}}]{Zezas2003}
{Zezas}, A., {Ward}, M.~J., \& {Murray}, S.~S. 2003, ApJ, 594, L31

\end{thebibliography}

\clearpage

\begin{appendix}
\section{Additional figures and tables }

\begin{figure}[h]
 \centering
  \includegraphics[width=1.015\linewidth]{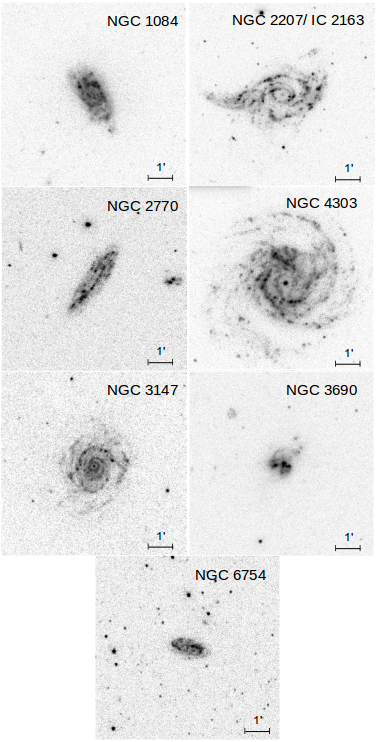}  
  \caption{Single epoch images taken with the \textit{Swift}/UVOT of the target galaxies: NGC\,1084, the system NGC\,2207/IC\,2163, NGC\,2770, NGC\,4303/M\,61, NGC\,3147, NGC\,3690, NGC\,6754. The angular size of each image is $7'\times7'$. }
  \label{pictures_UV}
 \end{figure}

\begin{figure}[h]
 \centering
  \includegraphics[width=1.\linewidth]{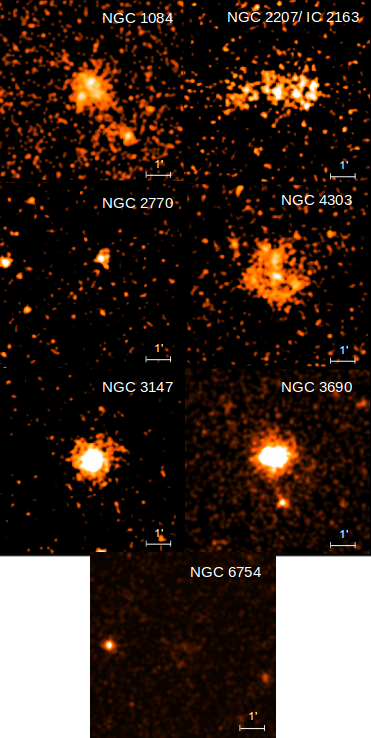}  
  \caption{Sum of all the images taken with the \textit{Swift}/XRT of the target galaxies: NGC\,1084, the system NGC\,2207/IC\,2163, NGC\,2770,  NGC\,4303/M\,61, NGC\,3147, NGC\,3690, NGC\,6754. The angular size of each image is $7'\times7'$.}
  \label{pictures_X}
 \end{figure}


\begin{table}[h]
\centering
\caption{Identification number, starting date, exposure time of observations carried out by the \textit{Swift} instruments XRT and UVOT during the monitoring of NGC\,1084.}
\footnotesize
\begin{tabular}{c c c c}
 \multicolumn{4}{c}{NGC\,1084}\\
 \hline
 \hline
 Obs ID & Obs Date & X--ray Exposure  & UV Exposure \\   
\hline
32999001 & 2013-11-01 & 1.5E+03\,s & - \\
32999002 & 2013-11-09 & 1.9E+03\,s & 1.8E+03\,s \\
32999003 & 2013-11-17 & 2.0E+03\,s & 2.0E+03\,s \\
32999004 & 2013-11-23 & 1.9E+03\,s & 1.9E+03\,s \\
32999005 & 2013-11-30 & 2.2E+03\,s & 2.2E+03\,s \\
32999006 & 2013-12-07 & 2.0E+03\,s & 2.0E+03\,s \\
32999007 & 2013-12-14 & 1.9E+03\,s & 1.9E+03\,s \\
32999008 & 2013-12-21 & 1.9E+03\,s & 1.9E+03\,s \\
32999009 & 2013-12-28 & 1.9E+03\,s & 1.9E+03\,s \\
32999010 & 2014-01-04 & 2.2E+03\,s & 2.2E+03\,s \\
32999011 & 2014-01-11 & 1.8E+03\,s & 1.8E+03\,s \\
32999012 & 2014-01-18 & 2.0E+03\,s & 2.0E+03\,s \\
32999013 & 2014-01-25 & 1.9E+03\,s & 1.9E+03\,s \\
32999014 & 2014-02-01 & 2.1E+03\,s & 2.1E+03\,s \\
32999015 & 2014-02-08 & 1.9E+03\,s & 1.9E+03\,s \\
32999016 & 2014-02-15 & 1.9E+03\,s & 1.8E+03\,s \\
32999017 & 2014-02-22 & 1.8E+03\,s & 1.8E+03\,s \\
32999018 & 2014-03-01 & 1.9E+03\,s & 1.8E+03\,s \\
32999019 & 2014-03-08 & 2.2E+03\,s & 2.2E+03\,s \\
32999020 & 2014-03-13 & 2.0E+03\,s & 1.1E+03\,s \\
\hline
\end{tabular}
\label{observations_NGC1084}
\end{table}


\begin{table}[h]
\centering
\caption{Identification number, starting date, exposure time of observations carried out by the \textit{Swift} instruments XRT and UVOT during the monitoring of the system NGC\,2207/IC\,2163.}
\footnotesize
\begin{tabular}{c c c c}
 
 \multicolumn{4}{c}{NGC\,2207/IC\,2163}\\
 \hline
 \hline
 Obs ID & Obs Date & X--ray Exposure  & UV Exposure \\   
\hline
33002001 & 2013-11-03 &1.6E+03\,s &  1.6E+03\,s \\
33002002 & 2013-11-10 &2.0E+03\,s & - \\ 
33002003 & 2013-11-17 &1.9E+03\,s & 1.9E+03\,s \\
33002004 & 2013-11-24 &1.9E+03\,s & 1.9E+03\,s \\
33002005 & 2013-12-01 &2.2E+03\,s & 2.2E+03\,s \\
33002007 &2013-12-11  &1.8E+03\,s &1.8E+03\,s \\
33002008 &2013-12-15  &2.2E+03\,s &2.2E+03\,s \\
33002009 &2013-12-22  &1.7E+03\,s &1.8E+03\,s \\
33002010 &2013-12-29  &1.4E+03\,s &1.3E+03\,s \\
33002011 &2014-01-05  &1.9E+03\,s &1.9E+03\,s \\
33002012 &2014-01-12  &2.0E+03\,s &1.9E+03\,s \\
33002013 &2014-01-19  &1.9E+03\,s &1.9E+03\,s \\
33002014 &2014-01-26  &2.0E+03\,s &2.0E+03\,s \\
33002015 &2014-02-02  &2.1E+03\,s &2.1E+03\,s \\
33002016 &2014-02-09  &1.9E+03\,s &1.9E+03\,s \\
33002017 &2014-02-16  &2.1E+03\,s &2.1E+03\,s \\
33002018 &2014-02-23  &1.3E+03\,s &1.3E+03\,s \\
33002019 &2014-02-26  &1.2E+03\,s &1.2E+03\,s \\
33002020 &2014-03-02  &2.0E+03\,s &2.0E+03\,s \\
33002021 &2014-03-09  &2.2E+03\,s &2.2E+03\,s \\
33002022 &2014-03-16  &1.8E+03\,s &1.8E+03\,s \\
33002023 &2014-03-23  &1.8E+03\,s &1.8E+03\,s \\
33002024 &2014-03-30  &2.0E+03\,s &2.0E+03\,s \\
33002025 &2014-04-06  &2.0E+03\,s &2.0E+03\,s \\
33002026 &2014-04-13  &1.9E+03\,s &1.9E+03\,s \\
33002027 &2014-04-20  &2.2E+03\,s &2.2E+03\,s \\
33002028 &2014-04-27  &2.0E+03\,s &2.0E+03\,s \\
\hline
\end{tabular}
\label{observations_NGC2207}
\end{table}


\begin{table}[h]
\centering
\caption{Identification number, starting date, exposure time of observations carried out by the \textit{Swift} instruments XRT and UVOT during the monitoring of NGC\,2770.}
\label{observations_NGC2770}
\footnotesize
\begin{tabular}{c c c c}
 \multicolumn{4}{c}{NGC\,2770}\\
 \hline
 \hline
 Obs ID & Obs Date & X--ray Exposure  & UV Exposure \\   
\hline
33000001 &2013-12-10 &2.0E+03\,s & 2.0E+03\,s \\
33000002 &2013-12-17 &2.0E+03\,s & 1.9E+03\,s\\
33000003 &2013-12-24 &2.0E+03\,s & 1.9E+03\,s \\
33000004 &2013-12-31 &2.0E+03\,s & 1.9E+03\,s \\
33000005 &2014-01-07 &1.9E+03\,s & 1.9E+03\,s \\
33000007 &2014-01-16 &1.9E+03\,s & 1.9E+03\,s \\
33000008 &2014-01-21 &2.2E+03\,s & 2.2E+03\,s \\
33000009 &2014-01-28 &1.8E+03\,s & 1.8E+03\,s \\
33000010 &2014-02-04 &1.5E+03\,s & 1.5E+03\,s \\
33000011 &2014-02-11 &2.0E+03\,s & - \\
33000012 &2014-02-18 &2.0E+03\,s & 1.9E+03\,s \\
33000013 &2014-02-25 &2.7E+03\,s & 2.7E+03\,s \\
33000014 &2014-03-04 &1.8E+03\,s & 1.8E+03\,s \\
33000015 &2014-03-11 &2.2E+03\,s & 2.2E+03\,s \\
33000016 &2014-03-18 &1.7E+03\,s & 1.7E+03\,s \\
33000017 &2014-03-25 &2.6E+03\,s & 2.6E+03\,s \\
33000018 &2014-04-01 &2.2E+03\,s & 2.2E+03\,s \\
33000019 &2014-04-07 &1.9E+03\,s & 2.1E+03\,s \\
33000020 &2014-04-15 &1.8E+03\,s & 1.8E+03\,s \\
33000021 &2014-04-22 &2.2E+03\,s & 2.2E+03\,s \\
33000023 &2014-05-02 &2.2E+03\,s & 2.2E+03\,s \\
33000024 &2014-05-07 &1.9E+03\,s & 1.9E+03\,s \\
33000025 &2014-05-13 &1.9E+03\,s & 1.8E+03\,s \\
33000026 &2014-05-20 &2.2E+03\,s & 2.2E+03\,s \\
33000027 &2014-05-27 &1.9E+03\,s & 1.8E+03\,s \\
33000028 &2014-06-04 & - & 2.1E+03\,s \\
\hline
\end{tabular}
\end{table}


\begin{table}[h]
\centering
\caption{Identification number, starting date, exposure time of observations carried out by the \textit{Swift} instruments XRT and UVOT during the monitoring of NGC\,4303.}
\label{observations_NGC4303}
\footnotesize
\begin{tabular}{c c c c}
 \multicolumn{4}{c}{NGC\,4303}\\
 \hline
 \hline
 Obs ID & Obs Date & X--ray Exposure  & UV Exposure \\   
\hline
33001001 &2014-01-24 &2.0E+03\,s &2.0E+03\,s\\
33001003 &2014-02-06 &1.6E+03\,s &1.6E+03\,s\\
33001004 &2014-02-13 &2.1E+03\,s &2.0E+03\,s\\
33001005 &2014-02-20 &2.0E+03\,s &1.9E+03\,s\\
33001006 &2014-02-27 &8.2E+02\,s &8.3E+02\,s\\
33001007 &2014-03-04 &1.9E+03\,s &1.9E+03\,s \\
33001008 &2014-03-06 &2.2E+03\,s & - \\
33001009 &2014-03-13 &1.7E+03\,s & - \\
33001010 &2014-03-20 &2.0E+03\,s &1.9E+03\,s\\
33001011 &2014-03-27 &2.0E+03\,s &2.0E+03\,s\\
33001012 &2014-03-27 &2.9E+03\,s &8.8E+02\,s\\
33001013 &2014-03-28 &4.7E+03\,s &1.1E+03\,s\\
33001014 &2014-04-03 &2.0E+03\,s &1.9E+03\,s\\
33001015 &2014-04-10 &1.5E+03\,s &1.4E+03\,s\\
33001016 &2014-04-17 &2.1E+03\,s &1.6E+03\,s\\
33001017 &2014-04-24 &1.9E+03\,s &1.9E+03\,s\\
33001018 &2014-05-01 &1.7E+03\,s &1.7E+03\,s\\
33001019 &2014-05-08 &2.2E+03\,s &2.1E+03\,s\\
33001020 &2014-05-15 &1.3E+03\,s &1.3E+03\,s\\
33001021 &2014-05-22 &2.0E+03\,s &2.0E+03\,s\\
33001022 &2014-05-29 &2.1E+03\,s &1.0E+03\,s\\
33001023 &2014-06-04 &    -       &1.9E+03\,s\\
\hline
\end{tabular}
\end{table}


\begin{table}[h]
\centering
\caption{Identification number, starting date, exposure time of observations carried out by the \textit{Swift} instruments XRT and UVOT during the monitoring of NGC\,3690.}
\label{observations_NGC3690}
\footnotesize
\begin{tabular}{c c c c}
 \multicolumn{4}{c}{NGC\,3690}\\
 \hline
 \hline
 Obs ID & Obs Date & X--ray Exposure  & UV Exposure \\   
\hline
32998001 &2014-01-10 &1.9E+03\,s &1.9E+03\,s \\
32998002 &2014-01-17 &1.9E+03\,s & - \\
32998003 &2014-01-24 &2.2E+03\,s&2.2E+03\,s \\
32998004 &2014-01-31 &1.9E+03\,s&1.9E+03\,s \\
32998005 &2014-02-07 &2.0E+03\,s&2.0E+03\,s \\
32998006 &2014-02-14 &1.4E+03\,s&1.4E+03\,s \\
32998007 &2014-02-21 &5.2E+03\,s& - \\
32998008 &2014-02-26 &9.5E+02\,s&9.5E+02\,s \\
32998009 &2014-02-28 &1.7E+03\,s&1.8E+03\,s \\
32998010 &2014-03-07 &2.1E+03\,s&2.1E+03\,s\\
32998011 &2014-03-14 &1.8E+03\,s&1.9E+03\,s\\
32998012 &2014-03-21 &1.7E+03\,s&1.6E+03\,s\\
32998013 &2014-03-28 &1.9E+03\,s&1.9E+03\,s\\
32998014 &2014-04-04 &2.1E+03\,s&2.1E+03\,s\\
32998015 &2014-04-11 &2.2E+03\,s&2.2E+03\,s\\
32998016 &2014-04-18 &1.7E+03\,s&1.6E+03\,s\\
32998017 &2014-04-25 &1.5E+03\,s&1.5E+03\,s\\
32998018 &2014-05-02 &1.5E+03\,s&1.5E+03\,s\\
32998020 &2014-05-16 &1.9E+03\,s&1.8E+03\,s\\
32998021 &2014-05-23 &2.0E+03\,s&2.0E+03\,s\\
32998022 &2014-05-30 &2.3E+03\,s&2.3E+03\,s\\
32998023 &2014-06-06 & - &2.0E+03\,s\\
32998024 &2014-06-13 &1.9E+03\,s&1.9E+03\,s\\
32998025 &2014-06-20 &1.9E+03\,s&1.9E+03\,s\\
32998026 &2014-06-27 &1.5E+03\,s&1.5E+03\,s\\
32998027 &2014-07-04 &1.8E+03\,s&1.8E+03\,s\\
32998028 &2014-07-11 &2.0E+03\,s&2.0E+03\,s\\
32998029 &2014-07-18 &2.2E+03\,s&2.2E+03\,s\\
32998030 &2014-07-25 &1.9E+03\,s&1.9E+03\,s\\
32998031 &2014-08-08 &2.1E+03\,s&2.1E+03\,s\\
32998032 &2014-08-15 &1.9E+03\,s&1.9E+03\,s\\
32998033 &2014-08-22 &2.7E+03\,s&2.0E+03\,s\\
32998034 &2014-08-29 &2.1E+03\,s&2.1E+03\,s\\
32998035 &2014-09-05 &1.8E+03\,s&1.8E+03\,s\\
32998037 &2014-09-18 &2.1E+03\,s&2.1E+03\,s \\
32998038 &2014-09-26 &1.9E+03\,s&1.9E+03\,s \\

\hline
\end{tabular}
\end{table}


\begin{table}[h]
\centering
\caption{Identification number, starting date, exposure time of observations carried out by the \textit{Swift} instruments XRT and UVOT during the monitoring of NGC\,6754.}
\label{observations_NGC6754}
\footnotesize
\begin{tabular}{c c c c}
 \multicolumn{4}{c}{NGC\,6754}\\
 \hline
 \hline
 Obs ID & Obs Date & X--ray Exposure  & UV Exposure \\   
\hline
33225001 &2014-04-06 &1.8E+03\,s&1.7E+03\\
33225002 &2014-04-13 &2.0E+03\,s&2.0E+03\\
33225003 &2014-04-20 &2.1E+03\,s&2.1E+03\\
33225004 &2014-04-27 &1.7E+03\,s&1.6E+03\\
33225005 &2014-05-04 &2.1E+03\,s&2.1E+03\\
33225006 &2014-05-11 &2.0E+03\,s&2.0E+03\\
33225007 &2014-05-18 &1.7E+03\,s&1.7E+03\\
33225008 &2014-05-25 &2.0E+03\,s&2.0E+03\\
33225009 &2014-06-01 &2.0E+03\,s&2.0E+03\\
33225011 &2014-06-15 &2.0E+03\,s&2.0E+03\\
33225012 &2014-06-22 &1.7E+03\,s&1.7E+03\\
33225013 &2014-06-29 &2.0E+03\,s&1.9E+03\\
33225014 &2014-07-06 &2.1E+03\,s&2.0E+03\\
33225015 &2014-07-13 &1.7E+03\,s&1.7E+03\\
33225016 &2014-07-17 &1.7E+03\,s&1.7E+03\\
33225017 &2014-07-20 &2.3E+03\,s&2.2E+03\\
33225018 &2014-07-27 &1.9E+03\,s&1.9E+03\\
33225019 &2014-08-03 &1.9E+03\,s&1.9E+03\\
33225020 &2014-08-10 &2.2E+03\,s&2.2E+03\\
33225021 &2014-08-17 &2.4E+03\,s&2.4E+03\\
33225022 &2014-08-24 &5.9E+02\,s&5.9E+02\\
33225023 &2014-08-31 &6.1E+02\,s&6.2E+02\\
33225024 &2014-09-04 &1.4E+03\,s&1.4E+03\\
33225025 &2014-09-07 &2.0E+03\,s&2.0E+03\\
33225026 &2014-09-14 &1.9E+03\,s&1.9E+03\\
33225027 &2014-09-21 &3.1E+03\,s&3.1E+03\\
33225028 &2014-09-28 &2.0E+03\,s&2.0E+03\\
\hline
\end{tabular}
\end{table}

\onecolumn

\begin{table}[h]
\centering
\caption{Identification number, starting date, exposure time of observations carried out by the \textit{Swift} instruments XRT and UVOT during the monitoring of NGC\,3147.
In the last column, the photon index of the absorbed power-law spectrum computed \citep[through the \textit{Swift}/XRT online tools;][]{Evans2009} in the XRT 0.3-10\,KeV with a Galactic column density 
of 3.30\,$\times\,10^{20}$\,cm$^{-2}$ \citep{Willingale2013}.}
\label{observations_NGC3147}
\footnotesize
\begin{tabular}{c c c c c}
 \multicolumn{5}{c}{NGC\,3147}\\
 \hline
 \hline
 Obs ID & Obs Date & X-ray Exposure  & UV Exposure &Index (0.3-10KeV)\\   
\hline
37992002 &2013-12-25 &1.8E+03\,s&1.8E+03\,s& 1.7 (+0.4, -0.4)  \\
37992003 &2014-01-01 &1.9E+03\,s&1.9E+03\,s& 1.31 (+0.45, -0.17)\\
37992004 &2014-01-15 &2.1E+03\,s&2.1E+03\,s& 1.25 (+0.34, -0.21)\\
37992005 &2014-01-22 &2.1E+03\,s&2.1E+03\,s& 1.17 (+0.34, -0.17)\\
37992006 &2014-01-29 &1.9E+03\,s&1.9E+03\,s& 1.30 (+0.39, -0.29)\\
37992007 &2014-02-05 &2.0E+03\,s&2.0E+03\,s& 1.7 (+0.5, -0.4)   \\
37992008 &2014-02-12 &1.8E+03\,s&1.7E+03\,s& 1.6 (+0.4, -0.4)   \\
37992009 &2014-02-19 &2.1E+03\,s&2.1E+03\,s& 1.49 (+0.31, -0.22)\\
37992010 &2014-03-06 &1.9E+03\,s&2.1E+03\,s& 1.48 (+0.36, -0.29)\\
37992011 &2014-03-12 &6.1E+02\,s&6.2E+02\,s& 1.5 (+0.8, -0.6)   \\
37992012 &2014-03-16 &1.4E+03\,s&1.4E+03\,s& 1.4 (+0.5, -0.3)   \\
37992013 &2014-03-19 &2.0E+03\,s&2.0E+03\,s& 1.48 (+0.43, -0.28)\\
37992014 &2014-03-26 &2.2E+03\,s&2.2E+03\,s& 1.47 (+0.39, -0.29)\\
37992015 &2014-04-02 &1.8E+03\,s&1.8E+03\,s& 1.66 (+0.48, -0.29)\\
37992016 &2014-04-09 &2.3E+03\,s&2.3E+03\,s& 1.52 (+0.39, -0.15)\\
37992017 &2014-04-16 &1.8E+03\,s&1.7E+03\,s& 1.38 (+0.34, -0.17)\\
37992018 &2014-04-26 &1.3E+03\,s&1.3E+03\,s& 1.7 (+0.5, -0.4)   \\
37992019 &2014-04-30 &1.7E+03\,s&1.7E+03\,s& 2.0 (+0.6, -0.5)   \\
37992020 &2014-05-07 &1.9E+03\,s&1.9E+03\,s& 1.8 (+0.4, -0.4)   \\
37992021 &2014-05-14 &2.0E+03\,s&1.9E+03\,s& 1.42 (+0.28, -0.23)\\
37992022 &2014-05-21 &2.2E+03\,s&2.2E+03\,s& 1.33 (+0.43, -0.22)\\
37992023 &2014-05-28 &2.1E+03\,s&2.1E+03\,s& 1.49 (+0.45, -0.26)\\					  
37992024 &2014-06-04 & -	 &2.2E+03\,s 		&-	\\
37992026 &2014-06-22 &2.1E+03\,s&2.1E+03\,s & 1.9 (+0.5, -0.5)   \\
37992027 &2014-06-25 &1.9E+03\,s&1.9E+03\,s & 1.42 (+0.51, -0.22)\\
37992028 &2014-07-02 &2.1E+03\,s&2.1E+03\,s & 1.6 (+0.5, -0.3)  \\
37992029 &2014-07-09 &1.3E+03\,s& -         & 1.5 (+0.5, -0.4)  \\
37992030 &2014-07-13 &1.0E+03\,s&1.0E+03\,s & 1.9 (+0.6, -0.5)  \\ 
\hline
\end{tabular}
\end{table}

\end{appendix}

\end{document}